\documentclass[journal]{IEEEtran}
\IEEEoverridecommandlockouts
\usepackage{amsmath,amsfonts}
\usepackage{tikz}
\usetikzlibrary{shapes.geometric, arrows.meta, positioning,decorations.pathreplacing, angles, quotes, calc}
\usepackage{mathtools}
\usepackage{amsthm}
\usepackage{algorithm}
\usepackage{algpseudocode}
\usepackage{bm}
\usepackage{amssymb}
 \usepackage{multirow}
\usepackage{array}
\usepackage{textcomp}
\usepackage{stfloats}
\usepackage{url}
\usepackage{microtype}
\usepackage{verbatim}
\usepackage{graphicx}
\usepackage{booktabs}
\usepackage{cite}
\usepackage[T1]{fontenc}
\usepackage[colorlinks=true, linkcolor=blue, citecolor=blue, urlcolor=blue]{hyperref}
\usepackage{enumitem}
\usepackage{fix-cm}
\DeclarePairedDelimiter\ceil{\lceil}{\rceil}

\usepackage{stfloats}
\usepackage{lipsum} 
\usepackage[font=small,skip=0pt]{caption}
\usepackage{subcaption}
\usepackage[export]{adjustbox}
\usepackage[inkscapelatex=false]{svg}

\theoremstyle{definition}

\usepackage{units}
\newcommand{\nth}[1]{{#1}^{\text{th}}}

\newcommand{\abs}[1]{\left|{#1}\right|}

\newcommand{\RD}[0]{r_{\mathrm{\scalebox{0.7} {RD} }}}
\newcommand{\etaangle}[0]{\eta_{\mathrm{\scalebox{0.7} {a} }}}
\newcommand{\etarange}[0]{\eta_{\mathrm{\scalebox{0.7} {r} }}}

\usepackage{standalone}
\usepackage{acronym}
\input{acronym}
\begin{document}
\title{RF Chain Count to Mitigate Beam Split in Wideband Hybrid Arrays: From Far-Field to Near-Field
\author{Ahmed Hussain, Asmaa Abdallah,~\IEEEmembership{Senior Member, IEEE}, Abdulkadir Celik,~\IEEEmembership{Senior Member, IEEE}, \\ and Ahmed M. Eltawil,~\IEEEmembership{Senior Member, IEEE}

\thanks{Ahmed Hussain, Asmaa Abdallah, and Ahmed M. Eltawil are with the Computer, Electrical, and Mathematical Sciences and Engineering (CEMSE) Division, King Abdullah University of Science and Technology (KAUST), Thuwal, 23955-6900, KSA. Abdulkadir Celik is with the School of Electronics and Computer Science, University of Southampton, SO17 1BJ, UK.}
}
}

\maketitle
\begin{abstract}
Deploying large antenna arrays transitions wireless communication into the radiative near-field (NF), where spherical wavefronts enable beamfocusing in the joint angle-distance domain. However, wideband beamforming with large antenna arrays suffers from beam split across both spatial dimensions. Although fully digital arrays can eliminate this impairment, their prohibitive hardware cost and power consumption motivate the adoption of hybrid array architectures equipped with a limited number of radio-frequency (RF) chains. This paper investigates the minimum number of RF chains required to mitigate the beam split. We derive closed-form expressions for the required number of RF chains in fully-connected (FC) architectures and propose an algorithm for sub-connected (SC) architectures, considering both uniform linear arrays (ULAs) and uniform circular arrays (UCAs). Our analysis reveals that beyond a focal distance of $10L$, where $L$ denotes the aperture length, the NF RF chain count converges to its far-field counterpart and becomes a function of the angle alone. Moreover, the FC hybrid architecture mitigates the beam split by using fewer RF chains than the SC architecture, albeit at the cost of increased complexity in the phase shifter network.
\end{abstract}


\section{Introduction} \label{sec-I}
\IEEEPARstart{A}{nalog} \acp{PS} apply a frequency-independent phase shift, whereas the propagation phase across an antenna array is frequency-dependent \cite{11050910}. Thus, an analog beamformer tuned to the carrier frequency becomes misaligned as the operating frequency changes, causing a loss of beamforming gain. In the \ac{FF}, this mismatch causes the beam to drift toward different angles across the signal bandwidth, a phenomenon known as \emph{beam squint} \cite{10934779}. The effect becomes more pronounced as signal bandwidth and array aperture increase, making beam squint a challenge in wideband \ac{mmWave} systems \cite{11050910,10220205}.

As the aperture becomes electrically large, communication transitions from the \ac{FF} to the radiative \ac{NF}. Consequently, conventional beamsteering evolves into \emph{beamfocusing}, where the array concentrates energy at a spatial location rather than merely along a direction \cite{10934779,10220205}. In this regime, the wideband counterpart of squint is \emph{beam split} \cite[Fig.~5]{9957130}, characterized by beam deviations in both the angle and distance dimensions. While \ac{TTD} beamforming can compensate for these impairments by introducing a frequency-dependent phase that tracks the propagation delay \cite{10541333,10458958}, the required delay ranges for large arrays are difficult to realize at \ac{mmWave} frequencies \cite{9833249}. Furthermore, analog \ac{TTD} networks incur substantial hardware complexity, power consumption, and cost \cite{9833249,10018114}. Hybrid architectures provide a practical alternative by operating between the fully analog and fully digital extremes, utilizing a limited number of \ac{RF} chains. The \ac{RF} chain count dictates the supported bandwidth for a given aperture and beamforming gain. This motivates the central question of this paper: \emph{What is the minimum number of \ac{RF} chains required by a wideband hybrid array to mitigate the \ac{NF} beam split?}

\subsection{Related Work} \label{subsec:related}
The spatial wideband effect manifests as inter-symbol interference in the time domain, loss of beamforming gain in the frequency domain, and the beam split in the spatial domain. Although a fully digital implementation can overcome these impairments through baseband processing, its hardware complexity scales directly with array size, rendering it impractical for massive arrays. Existing low-complexity architectures instead approximate the fully digital ideal through bandwidth-slicing or subarray-based designs \cite{9398864,lu2026wideband,10934783,631557,hussain2025near}. Bandwidth slicing partitions the operational bandwidth into narrow subbands, each with an independent phase-shift vector, because beam squint is negligible over sufficiently narrow frequency intervals \cite{9398864}. Alternatively, the array itself is partitioned into subarrays equipped with \acp{PS}, as shown in Fig.~\ref{fig:subarray_types}. In the simplest configuration, \acp{PS} are employed at the subarray level, which limits the array's scanning capability. The \ac{DPP} architecture \cite{10541333,9957130,10934783} improves upon this design by placing \acp{PS} at the element level and \acp{TTD} at the subarray level, which mitigates the scan limitation while containing the cost, since the \acp{TTD} are not deployed per element. Nevertheless, as highlighted earlier, large bandwidths and apertures require a wide range of delay values, which renders the \ac{TTD} design itself challenging. Hybrid beamforming architectures, encompassing both \ac{FC} and \ac{SC} configurations, leverage \ac{RF} chains to mitigate spatial wideband effects. While the minimum \ac{RF} chain requirement to suppress \ac{FF} beam squint in \ac{FC} arrays was derived in \cite{10495774}, that framework cannot be directly applied to the \ac{NF}, where beam split manifests across both angular and radial dimensions.

 The array geometry governs the beamfocusing characteristics of \ac{NF} arrays. Unlike the \ac{FF}, the \ac{NF} beams possess a finite beamdepth that enables both angle and range estimation \cite{wei2025fundamental}. For a fixed aperture length, the \ac{UCA} provides a narrower beamdepth and a larger effective \ac{NF} region, quantified by the \ac{EBRD}, than the \ac{ULA} \cite{11498324}. Moreover, \acp{UCA} achieve superior wideband sensing performance compared with \acp{ULA} of the same aperture \cite{10988576}. These observations suggest that the array geometry may also influence the \ac{RF}-chain requirement for beam split mitigation, motivating a comparison between \acp{ULA} and \acp{UCA}.

The operational bandwidth specifies the frequency range over which the system operates. In contrast, the \ac{IBW} denotes the frequency interval over which an antenna array preserves a prescribed beamforming gain \cite{haupt2015timed,brown2021active}. Conceptually, the \ac{IBW} is the frequency-domain analogue of beamwidth and beamdepth, which define the angle and range intervals, respectively, over which the beamforming gain exceeds a specified threshold. In the \ac{FF}, the \ac{IBW} metric is utilized to determine the number of \ac{RF} chains specifically for \ac{SC} architectures \cite{brown2021active}. However, the corresponding \ac{NF} characterization remains unavailable, as the dependence of the \ac{IBW} on both angle and range has not yet been established. This motivates the derivation of analytical expressions for the \ac{NF} \ac{IBW}, together with closed-form design rules for the required \ac{RF} chain count.

\subsection{Main Contributions}
To address the research gaps highlighted in Section \ref{subsec:related}, the main contributions of this paper are summarized as follows.
\begin{itemize}
\item \textbf{NF radial beam split boundary}: We adopt a hybrid beamforming architecture and characterize the frequency-dependent beamforming gain in the distance domain. We decouple \ac{NF} beam split into an angular component, which coincides with \ac{FF} beam squint, and a radial component exclusive to the \ac{NF}. Furthermore, we derive the boundary beyond which the radial beam split vanishes.

\item \textbf{RF chain count for SC hybrid arrays}: We derive closed-form \ac{NF} \ac{IBW} expressions for the \ac{ULA} and the \ac{UCA}. By applying the \ac{IBW} of the binding subarray, we propose an algorithm that determines the required \ac{RF} chain count for \ac{SC} hybrid architectures.

\item \textbf{RF chain count for FC hybrid arrays}: To determine the required \ac{RF} chain count for \ac{FC} hybrid arrays in the \ac{NF}, we analyze the eigenspectrum of the space-frequency correlation matrix. The number of dominant eigenvalues simplifies to the product of the bandwidth and the delay spread across the aperture, yielding the \ac{RF} chain requirements for both the \ac{ULA} and \ac{UCA}.

\item \textbf{NF vs. FF RF chain comparison}: Numerical results demonstrate that the \ac{NF} \ac{RF} chain requirement becomes range-invariant beyond a focal distance of $10L$ \footnote{Boresight is an exception for very large bandwidth.}, where $L$ is the aperture length, coinciding with the \ac{FF} \ac{RF} chain count. Furthermore, \ac{FC} architectures require fewer \ac{RF} chains than \ac{SC} architectures, but at the cost of a more complex phase shifter network.
\end{itemize}

\subsection{Notation and Organization of the Paper}
Matrices, vectors, and scalars are denoted by bold uppercase, bold lowercase, and lowercase letters, respectively. The transpose and conjugate transpose are denoted by $(\cdot)^{\mathrm{T}}$ and $(\cdot)^{{\mathsf{H}}}$, respectively. The Euclidean ($\ell_2$) norm of a vector is denoted by $\|\cdot\|$, the absolute value of a scalar by $|\cdot|$, the ceiling operator by $\lceil\cdot\rceil$, and the trace of a square matrix by $\operatorname{tr}(\cdot)$. 

The remainder of this paper is organized as follows. Section~\ref{sec-II} presents the system model. Section~\ref{sec-III} characterizes the beam split and \ac{IBW} for the \ac{ULA}. Section~\ref{sec-IV} derives the minimum \ac{RF} chain count to mitigate beam split across both \ac{SC} and \ac{FC} architectures for \ac{ULA}. Section~\ref{sec-V} extends this framework to the \ac{UCA}. Section~\ref{sec-VI} presents numerical results to validate the analysis, and Section~\ref{sec-VII} concludes the paper.

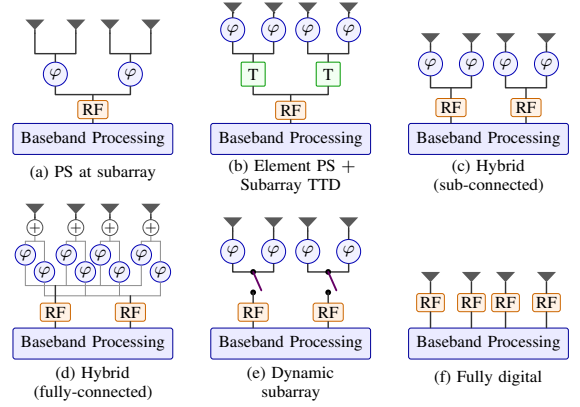
\begin{figure}[t]
\centering
\resizebox{.9\columnwidth}{!}{\begin{tikzpicture}[
  x=1cm, y=1cm, line cap=round, line join=round,
  font=\scriptsize\rmfamily,
  wire/.style={draw=black!80, line width=0.6pt},
  thin wire/.style={draw=black!45, line width=0.35pt},
  ps/.style={circle, draw=blue!70!black, fill=blue!5, inner sep=0pt, minimum size=10pt},
  ttd/.style={rectangle, rounded corners=0.5pt, draw=green!60!black, fill=green!5, inner sep=1pt, minimum size=10pt},
  rf/.style={rectangle, rounded corners=1pt, draw=orange!80!black, fill=orange!10, inner xsep=2pt, inner ysep=1.5pt},
  bb/.style={rectangle, rounded corners=1pt, draw=blue!60!black, fill=blue!8, minimum width=2.3cm, minimum height=0.35cm},
  add/.style={circle, draw=black!70, fill=white, inner sep=0pt, minimum size=7.5pt},
  sw line/.style={draw=violet!80!black, line width=0.8pt},
  ttl/.style={font=\scriptsize\rmfamily, text width=2.8cm, align=center}
]

\def\xa{-0.85} \def\xb{-0.25} \def\xc{0.25} \def\xd{0.85}
\def\xL{-0.55} \def\xR{0.55}
\def\PITCH{2.9}
\def\YSHIFT{3.0}

\newcommand{\ANT}[1]{%
  \fill[black!65] (#1-0.14, \yT) -- (#1+0.14, \yT) -- (#1, \yB) -- cycle;
  \draw[wire] (#1, \yB) -- (#1, \yF);
}

\begin{scope}[shift={(0,\YSHIFT)}]
  \def\yT{1.35} \def\yB{1.2} \def\yF{1.15}
  \foreach \x in {\xa,\xb,\xc,\xd} { \ANT{\x} }
  
  \draw[wire] (\xa, \yF) -- (\xa, 0.85) -- (\xb, 0.85) -- (\xb, \yF);
  \draw[wire] (\xc, \yF) -- (\xc, 0.85) -- (\xd, 0.85) -- (\xd, \yF);
  
  \node[ps] (p1) at (\xL, 0.55) {$\varphi$};
  \node[ps] (p2) at (\xR, 0.55) {$\varphi$};
  \draw[wire] (\xL, 0.85) -- (p1.north);
  \draw[wire] (\xR, 0.85) -- (p2.north);
  
  \draw[wire] (p1.south) -- (\xL, 0.25) -- (\xR, 0.25) -- (p2.south);
  \node[rf] (rf) at (0, 0.03) {RF};
  \draw[wire] (0, 0.25) -- (rf.north);
  
  \node[bb] (bb) at (0, -0.4) {Baseband Processing};
  \draw[wire] (rf.south) -- (bb.north);
  \node[ttl] at (0, -0.9) {(a) PS at subarray};
\end{scope}

\begin{scope}[shift={(\PITCH,\YSHIFT)}]
  \def\yT{1.60} \def\yB{1.45} \def\yF{1.45}
  \foreach \i/\x in {1/\xa, 2/\xb, 3/\xc, 4/\xd} {
      \ANT{\x}
      \node[ps] (p\i) at (\x, 1.20) {$\varphi$};
      \draw[wire] (\x, \yF) -- (p\i.north);
  }
  
  \draw[wire] (p1.south) -- (\xa, 0.85) -- (\xb, 0.85) -- (p2.south);
  \draw[wire] (p3.south) -- (\xc, 0.85) -- (\xd, 0.85) -- (p4.south);
  
  \node[ttd] (tL) at (\xL, 0.55) {T};
  \node[ttd] (tR) at (\xR, 0.55) {T};
  \draw[wire] (\xL, 0.85) -- (tL.north);
  \draw[wire] (\xR, 0.85) -- (tR.north);
  
  \draw[wire] (tL.south) -- (\xL, 0.25) -- (\xR, 0.25) -- (tR.south);
  \node[rf] (rf) at (0, 0.02) {RF};
  \draw[wire] (0, 0.25) -- (rf.north);
  
  \node[bb] (bb) at (0, -0.4) {Baseband Processing};
  \draw[wire] (rf.south) -- (bb.north);
  \node[ttl] at (0, -0.95) {(b) Element PS $+$ \\ Subarray TTD};
\end{scope}

\begin{scope}[shift={(2*\PITCH,\YSHIFT)}]
  \def\yT{1.15} \def\yB{1.0} \def\yF{.95}
  \foreach \i/\x in {1/\xa, 2/\xb, 3/\xc, 4/\xd} {
      \ANT{\x}
      \node[ps] (p\i) at (\x, 0.70) {$\varphi$};
      \draw[wire] (\x, \yF) -- (p\i.north);
  }
  
  \draw[wire] (p1.south) -- (\xa, 0.35) -- (\xb, 0.35) -- (p2.south);
  \draw[wire] (p3.south) -- (\xc, 0.35) -- (\xd, 0.35) -- (p4.south);
  
  \node[rf] (rfL) at (\xL, 0.06) {RF};
  \node[rf] (rfR) at (\xR, 0.06) {RF};
  \draw[wire] (\xL, 0.35) -- (rfL.north);
  \draw[wire] (\xR, 0.35) -- (rfR.north);
  
  \node[bb] (bb) at (0, -0.4) {Baseband Processing};
  \draw[wire] (rfL.south) -- (rfL.south |- bb.north);
  \draw[wire] (rfR.south) -- (rfR.south |- bb.north);
  \node[ttl] at (0, -0.95) {(c) Hybrid \\ (sub-connected)};
\end{scope}

\begin{scope}[shift={(0,0)}]
  \def\yT{1.65} \def\yB{1.5} \def\yF{1.45}
  \foreach \i/\x in {1/\xa, 2/\xb, 3/\xc, 4/\xd} {
      \ANT{\x}
      \node[
    add,
    minimum size=3pt,
    inner sep=0pt,
    font=\tiny
] (a\i) at (\x,1.30) {$+$};
      \draw[wire] (\x, \yF) -- (a\i.north);
      
      \node[ps, minimum size=8.5pt] (pl\i) at (\x-0.14, 0.9) {$\varphi$};
      \node[ps, minimum size=8.5pt] (pr\i) at (\x+0.14, 0.65) {$\varphi$};
      
      \draw[thin wire] (a\i.south) -- (\x, 1.10);
      \draw[thin wire] (\x-0.14, 1.10) -- (pl\i.north);
      \draw[thin wire] (\x+0.14, 1.10) -- (pr\i.north);
      \draw[thin wire] (\x-0.14, 1.10) -- (\x+0.14, 1.10);
  }
  
  \draw[thin wire] (\xa-0.14, 0.45) -- (\xd-0.14, 0.45);
  \draw[thin wire] (\xa+0.14, 0.3) -- (\xd+0.14, 0.3);
  
  \foreach \i/\x in {1/\xa, 2/\xb, 3/\xc, 4/\xd} {
      \draw[thin wire] (pl\i.south) -- (\x-0.14, 0.45);
      \draw[thin wire] (pr\i.south) -- (\x+0.14, 0.3);
  }
  
  \node[rf] (rfL) at (\xL, 0.06) {RF};
  \node[rf] (rfR) at (\xR, 0.06) {RF};
  \draw[wire] (\xL, 0.45) -- (rfL.north);
  \draw[wire] (\xR, 0.3) -- (rfR.north);
  
  \node[bb] (bb) at (0, -0.4) {Baseband Processing};
  \draw[wire] (rfL.south) -- (rfL.south |- bb.north);
  \draw[wire] (rfR.south) -- (rfR.south |- bb.north);
  \node[ttl] at (0, -0.99) {(d) Hybrid \\ (fully-connected)};
\end{scope}

\begin{scope}[shift={(\PITCH,0)}]
  \def\yT{1.45} \def\yB{1.3} \def\yF{1.25}
  \foreach \i/\x in {1/\xa, 2/\xb, 3/\xc, 4/\xd} {
      \ANT{\x}
      \node[ps] (p\i) at (\x, 0.95) {$\varphi$};
      \draw[wire] (\x, \yF) -- (p\i.north);
  }
  
  \draw[wire] (p1.south) -- (\xa, 0.65) -- (\xb, 0.65) -- (p2.south);
  \draw[wire] (p3.south) -- (\xc, 0.65) -- (\xd, 0.65) -- (p4.south);
  
  \fill (\xL, 0.65) circle (1.2pt);
  \fill (\xL, 0.35) circle (1.2pt);
  \draw[sw line] (\xL, 0.65) -- (\xL+0.12, 0.38);
  
  \fill (\xR, 0.65) circle (1.2pt);
  \fill (\xR, 0.35) circle (1.2pt);
  \draw[sw line] (\xR, 0.65) -- (\xR+0.12, 0.38);
  
  \node[rf] (rfL) at (\xL, 0.06) {RF};
  \node[rf] (rfR) at (\xR, 0.06) {RF};
  \draw[wire] (\xL, 0.35) -- (rfL.north);
  \draw[wire] (\xR, 0.35) -- (rfR.north);
  
  \node[bb] (bb) at (0, -0.4) {Baseband Processing};
  \draw[wire] (rfL.south) -- (rfL.south |- bb.north);
  \draw[wire] (rfR.south) -- (rfR.south |- bb.north);
  \node[ttl] at (0, -0.99) {(e) Dynamic \\ subarray};
\end{scope}

\begin{scope}[shift={(2*\PITCH,0)}]
  \def\yT{0.65} \def\yB{0.5} \def\yF{0.35}
  \foreach \i/\x in {1/\xa, 2/\xb, 3/\xc, 4/\xd} {
      \ANT{\x}
      \node[rf, inner xsep=1.5pt] (r\i) at (\x, 0.2) {RF};
      \draw[wire] (\x, \yF) -- (r\i.north);
  }
  
  \node[bb] (bb) at (0, -0.4) {Baseband Processing};
  \foreach \i/\x in {1/\xa, 2/\xb, 3/\xc, 4/\xd} {
      \draw[wire] (r\i.south) -- (r\i.south |- bb.north);
  }
  \node[ttl] at (0, -0.9) {(f) Fully digital};
\end{scope}

\end{tikzpicture}}
\caption{Categories of wideband beamforming architectures.}
\label{fig:subarray_types}
\end{figure}

\section{System Model} \label{sec-II}
We consider a downlink communication system where the \ac{BS} employs a large $N$-element antenna array oriented along the $x$-axis. We adopt a free-space \ac{LoS} \ac{NF} channel model. Denoting the carrier frequency by $f_c$ and the frequency offset relative to it by $f\in[-\tfrac{W}{2},\tfrac{W}{2}]$, with $W$ the bandwidth, the $\nth{n}$ entry of the \ac{NF} array response vector $\mathbf{a}(f) \in \mathbb{C}^{N} $ for a \ac{UE} located at a distance $r$ with respect to the array center is
\begin{equation}
[\mathbf{a}(f)]_n=\exp\!\Big(-2\pi j\frac{f_c+f}{c}\,\big(d(x_n)-r\big)\Big),
\label{eq:steer}
\end{equation}
where $d(x_n)$ denotes the distance from the \ac{UE} to the $\nth{n}$ element of the antenna array. The amplitude factor omitted in \eqref{eq:steer} is absorbed into the \ac{SNR} \cite{10495774}.
Furthermore, amplitude variation across different antenna elements is assumed negligible. 

We consider a hybrid beamforming architecture with $K$ \ac{RF} chains. The \ac{BS} employs a frequency-flat analog beamformer $\mathbf{W}_a\in\mathbb{C}^{N\times K}$ followed by a frequency-dependent digital beamformer $\mathbf{w}_d(f)\in\mathbb{C}^{K}$. The received signal
corresponding to the transmit symbol $q(f)$ is
\begin{equation}
y(f)=\mathbf{a}^{\mathsf{H}}(f)\mathbf{W}_a\mathbf{w}_d(f)\,q(f)+v(f),
\label{eq:received_signal}
\end{equation}
where $v(f)$ is additive white Gaussian noise with unit variance. The \emph{beamforming gain} at frequency $f$ is
\begin{equation}
g(f)=\frac{\big|\mathbf{a}^{\mathsf{H}}(f)\mathbf{W}_a\mathbf{w}_d(f)\big|^2}
{\big\|\mathbf{W}_a\mathbf{w}_d(f)\big\|^2}.
\label{eq:gf}
\end{equation}
For a fixed analog stage $\mathbf{W}_a$, we can find the optimal digital beamformer using the \ac{LS} estimator as \cite{7448873}
\begin{equation}
\mathbf{w}_d(f) = (\mathbf{W}_a^{\mathsf{H}}\mathbf{W}_a)^{-1}\mathbf{W}_a^{\mathsf{H}}\mathbf{a}(f),
\label{eq:gf_quot}
\end{equation}
which bounds the maximum beamforming gain as
\begin{equation}
g(f) = \big\|(\mathbf{W}_a^{\mathsf{H}}\mathbf{W}_a)^{-1/2}\mathbf{W}_a^{\mathsf{H}}\mathbf{a}(f)\big\|^2.
\label{eq:gf_bound}
\end{equation}
Furthermore, the average beamforming gain is defined as 
\begin{equation}
g_{\mathrm{av}}=\frac{1}{W}\int_{-W/2}^{W/2}g(f)\,df,
\label{eq:gavg} 
\end{equation}
which is averaged over the bandwidth $W$.

\section{Beam Split and Instantaneous Bandwidth} \label{sec-III}
We consider a \ac{BS} equipped with a \ac{ULA} with inter-element spacing ($\Delta=\lambda/2$) and aperture length\footnote{The aperture length refers to the largest physical dimension of the array. Specifically, it corresponds to the total length for a \ac{ULA} and the diameter for a \ac{UCA}.} $L \approx N\Delta$. A single-antenna \ac{UE} is located at a distance $r$ from the array center and an angle $\varphi$ measured with respect to the array normal. The position of the $\nth{n}$ antenna element is given by $x_n = \left(n-\frac{N}{2}\right)\Delta$, and the corresponding element--\ac{UE} distance is
\begin{subequations}
\begin{align}
d(x_n) &=\sqrt{r^2-2rx_n\sin\varphi+x_n^2} \label{eq:fresnel_exact} \\
&\overset{(a)}{\approx}r-x_n\sin\varphi+\frac{x_n^2\cos^2\varphi}{2r}, 
\label{eq:fresnel}
\end{align}
\end{subequations}
where $(a)$ follows from the Fresnel approximation \cite{10988573,11428208}. The \ac{FF} corresponds to $r\to\infty$, where the quadratic term vanishes, and the wavefront becomes planar. The classical boundary between \ac{NF} and \ac{FF} is given by the Rayleigh distance $\RD = \frac{2L^2}{\lambda}$ \cite{11095387}. In \eqref{eq:fresnel}, the linear term determines the beam direction, while the quadratic term encodes the wavefront curvature associated with the focusing distance.

Beam split in the \ac{NF} across both angle and distance can be observed from the beamforming gain $g(f)$ in \eqref{eq:gf} on the space--frequency plane. To expose the beam split, we omit the digital stage and fix the analog beamformer with a single \ac{RF} chain $\mathbf{w}_a$ at the carrier frequency, which reduces \eqref{eq:gf} to
\begin{equation}
g(f)=\big|\mathbf{w}^{\mathsf{H}}_a\mathbf{a}(f)\big|^2 .
\label{eq:g_analog}
\end{equation}
Let the \ac{NF} beam be focused towards $(\varphi_0,r_0)$ at the carrier frequency $f_c$. Combining \eqref{eq:steer}--\eqref{eq:fresnel} with \eqref{eq:g_analog}, the gain becomes
\begin{equation}
g(f)=\left|\sum_{n=1}^{N}
\exp\!\left(-\,j\frac{2\pi}{c}\,\Xi_n(f)\right)\right|^2,
\label{eqn12}
\end{equation}
where the per-element phase mismatch is
\begin{equation}
\begin{split}
\Xi_n(f) &= \underbrace{\frac{x^2_n}{2}\!\left(f_c\frac{\cos^2\varphi_0}{r_0}
-(f_c+f)\frac{\cos^2\varphi}{r}\right)}_{\text{curvature (distance) term}} \\
&\quad - \underbrace{x_n\big(f_c\sin\varphi_0-(f_c+f)\sin\varphi\big)}_{\text{tilt (angle) term}} .
\end{split}
\label{eq:Xi}
\end{equation}
To evaluate the beam split, the coefficients of the linear and quadratic terms in \eqref{eq:Xi} are set to zero. Each condition is subsequently analyzed in turn.

\subsection{Angular Beam Split}
For the frequency-dependent propagation phase to perfectly match the fixed analog phase, the tilt term in \eqref{eq:Xi} requires
\begin{equation}
(f_c+f)\sin\varphi=f_c\sin\varphi_0
 \ \Longrightarrow \ 
\sin\varphi=\frac{1}{(1+\bar{f})}\,\sin\varphi_0 ,
\label{eq:squint_angle}
\end{equation}
where $\bar{f} = \frac{f}{f_c}$ is normalized frequency offset. The beam, therefore, points at a different angle at every frequency, other than the carrier. Since $f\in[-\tfrac{W}{2},+\tfrac{W}{2}]$, the peak drifts \emph{toward} boresight for $f>0$ and \emph{away} from boresight for $f<0$. This frequency-dependent angular drift is referred to as \emph{angular beam split},\textbf{ which vanishes only at boresight ($\varphi_0=0$) and becomes more pronounced as the beam is steered toward larger angles}.

\subsection{Radial Beam Split}
Similarly, matching the frequency-dependent term to the fixed analog phase in the curvature component of \eqref{eq:Xi} requires
\begin{equation}
(f_c+f)\,\frac{\cos^2\varphi}{r}=f_c\,\frac{\cos^2\varphi_0}{r_0},
 \implies 
r= r_0 \, ({1+\bar{f}}) \,\frac{\cos^2\varphi}{\cos^2\varphi_0}.
\label{eq:split_match}
\end{equation}
By substituting \eqref{eq:squint_angle} into \eqref{eq:split_match}, the effect of angular beam split is removed, and \eqref{eq:split_match} simplifies to
\begin{equation}
r=\xi(\bar f,\varphi_0) r_0,
\qquad
\xi(\bar f,\varphi_0)=\frac{(1+\bar f)^2-\sin^2\varphi_0}
{(1+\bar f)\cos^2\varphi_0}.
\label{eq:split_range}
\end{equation}
This frequency-dependent focal-range drift is \emph{radial beam split}. The focal range drifts \emph{outward} ($r>r_0$) for $f>0$ and \emph{inward} ($r<r_0$)
for $f<0$. 

\subsection{NF Radial Beam Split Boundary}
The angular beam split vanishes at boresight, whereas the radial beam split is confined to the \ac{NF}. We use the beamdepth expression to determine the boundary within the \ac{NF} beyond which the radial beam split becomes negligible.

\subsubsection{Beamdepth at $f_c+f$}\label{sec-IIIC1}
The beamdepth is defined as the distance interval over which the array gain remains within $3$\,dB of its peak value \cite{10443535}. For a beam focused at distance $r_0$ at the carrier frequency $f_c$, the beamdepth is given by \cite{10934779}
\begin{equation} 
\mathrm{BD} = \frac{8\,\etarange^2 \,r_0^{2}\,\RD\cos^2\varphi_0}{(\RD \cos^2\varphi_0)^2-(4\,\etarange^2\,r_0)^2},
\label{eq:bd}
\end{equation}
where $\etarange = 1.318$, assuming a 3 dB loss threshold, $\rho_\epsilon=0.5$. Substituting the drifted range
\eqref{eq:split_range} for $r_0$, the squint angle \eqref{eq:squint_angle}
for $\varphi_0$, and the Rayleigh distance at the offset frequency
$(1+\bar f)\RD$ for $\RD$, the beamdepth at $f_c+f$ becomes
\begin{equation} 
\overline{\mathrm{BD}} = \xi(\bar f,\varphi_0)\, \mathrm{BD}.
\label{eq:bd_split}
\end{equation}
Beam split shifts the focal range and scales the beamdepth by the same factor. At boresight $\xi=1+\bar f$, and the beamdepth widens by
$(1+\bar f)$ for $f>0$ and narrows by $(1-|\bar f|)$ for $f<0$. Away from boresight $\xi$ carries the additional factor $\cos^2\varphi/\cos^2\varphi_0$, which exceeds unity for $f>0$. The monotonicity of $\xi$ in $\bar f$ makes this behavior consistent at all steering angles, with the beamdepth increasing as the focus shifts away from the array.

\begin{figure}[!t]
\centering
\includegraphics[width=0.9\columnwidth]{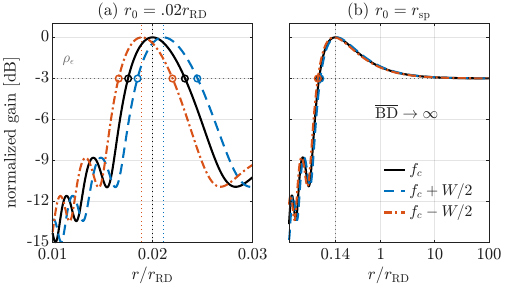}
\caption{Distance-domain gain at $f_c$ and $f_c\pm W/2$ for $r_0=0.02\RD$ in
(a) and $r_0=r_{\mathrm{sp}}=0.14\RD$ in (b). We set $f_c= \unit[28]{GHz}$, $N= 256$ and $W=\unit[3]{GHz}$. }
\label{fig:bd_ebrd}
\end{figure}

\subsubsection{Radial Beam Split Boundary}
The \ac{NF} beam split boundary, denoted as $r_{\mathrm{sp}}$, is the distance beyond which the radial beam split ceases to act in the \ac{NF}, leaving angular beam split as the sole impairment. It is obtained by setting the denominator of the beamdepth \eqref{eq:bd} to zero, and is given by
\begin{equation}
r_{\mathrm{sp}}=\frac{{\RD}\cos^2\varphi_0}{4\,\etarange^2}.
\label{eq:nfbb}
\end{equation}
\begin{proof}
As the focusing range $r_0$ increases, the denominator of the beamdepth expression \eqref{eq:bd_split}, which is the same as that in \eqref{eq:bd}, shrinks and the beamdepth increases. The boundary at which the beamdepth becomes infinite marks the range beyond which the focal drift can no longer be resolved, so that the radial beam split no longer reduces the gain. Setting the denominator of \eqref{eq:bd} to zero
\begin{equation}
(\RD\cos^2\varphi_0)^2-(4\,\etarange^2\,r_0)^2=0,
\label{eq:nfbb_denom}
\end{equation}
and solving for the positive root yields \eqref{eq:nfbb}. 
\end{proof}
Radial beam split in \eqref{eq:split_range} is most severe at short distances where the wavefront curvature across the aperture is pronounced. \textbf{It weakens with range, and vanishes at the beam split boundary $r_{\mathrm{sp}}$ in \eqref{eq:nfbb}}. This boundary coincides with the \ac{EBRD} boundary \cite{11428208,10443535}, which defines the maximum beamfocusing limit. Beyond this limit, the array can no longer resolve in the distance dimension, and the radial beam split disappears. Only the angular beam split remains for $r_0 > r_{\mathrm{sp}}$. To exemplify this, Fig.~\ref{fig:bd_ebrd} presents the distance-domain gain of a beam focused at boresight at $f_c$ and evaluated at the two band edges. At $r_0=0.02\,\RD$, the beamdepth for the upper band edge ($f_c+W/2$) shifts to the right and widens, while for the lower band edge ($f_c-W/2$), it shifts to the left and contracts. At $r_0=r_{\mathrm{sp}}=0.14\,\RD$, the right-side half-power boundary is absent, and the gain decays monotonically toward $\rho_\epsilon$ without crossing it. Consequently, beams have infinite depth at this range, as predicted by \eqref{eq:nfbb}. Finally, we highlight that the angular beam split is the same in the \ac{FF} and the \ac{NF}. This behavior mirrors the beamwidth, which also remains the same in the \ac{NF} and the \ac{FF} \cite{11428208}.

\subsection{Instantaneous Bandwidth}\label{subsec:ula_ibw}
In the \ac{NF}, angular and radial beam split drift the focal point, according to \eqref{eq:squint_angle} and \eqref{eq:split_range}, respectively. Both effects scale with the offset $f$ from the carrier frequency $f_c$ and reduce the beamforming gain at the focal point. The \ac{IBW} is then defined as the frequency deviation $f$ from the carrier frequency at which this gain reduces to an acceptable threshold $\rho_\epsilon$ \cite{9020885, haupt2015timed, brown2021active}. In electromagnetics, a 3dB threshold $\rho_\epsilon = 0.5$ is regarded as an acceptable threshold, most commonly used for beamwidth and beamdepth \cite{hansen2009phased,10988573,10934779,11428208,10443535}. 

To isolate the array-gain degradation caused solely by frequency offset in \eqref{eq:Xi}, we evaluate the phase mismatch at the beamfocusing point by setting $\varphi=\varphi_0$, $r=r_0$. Under these conditions, the per-element phase mismatch simplifies to $\Xi_n=- f\,\delta_n$, where $\delta_n=\frac{x^2_n\cos^2\varphi_0}{2r_0}-x_n\sin\varphi_0$. The normalized form of the gain expression in \eqref{eqn12} is then given by
\begin{equation}\small
g(f)=\left| \frac{1}{N} \sum_{n=1}^{N}
\exp\!\left(j\,\frac{2\pi f}{c}\,\delta_n\right)\right|^2,
\label{eqn-ibw-master}
\end{equation}
\normalsize
whose tilt term $x_n\sin\varphi_0$ and curvature term $\tfrac{x^2_n\cos^2\varphi_0}{2r_0}$ determine the angle-domain \ac{IBW} and the distance-domain \ac{IBW}, respectively. 
\subsubsection{Angle-Domain IBW}
Retaining the tilt term $x_n\sin\varphi_0$ of \eqref{eqn-ibw-master}, the angle domain gain $g_a(f)$ is given by
\begin{align}\small
g_a(f)
&=\left|\frac{1}{N}\sum_{n=1}^{N}
\exp\!\left(-\,j\,\frac{2\pi f}{c}\,x_n\sin\varphi_0\right)\right|^2 ,\\
&=\left|\frac{1}{N}\,
\frac{\sin\!\big(\pi N \Delta\, f\sin\varphi_0/c\big)}
{\sin\!\big(\pi \Delta\, f\sin\varphi_0/c\big)}\right|^2 .
\label{eqn-af-closed}
\end{align}
\normalsize
For large $N$ the denominator argument stays small across the mainlobe and \eqref{eqn-af-closed} reduces to $\operatorname{sinc}^2 (\etaangle)$ and
$\etaangle=\pi N\Delta f\sin\varphi_0/c$. To find \ac{IBW} at a specific threshold $\rho_\epsilon$, we set $g_a(\etaangle) = \rho_\epsilon$, which evaluates to $\etaangle \approx 1.391$ for $\rho_\epsilon = 0.5$. We can solve for one-sided frequency offset based on the $\etaangle$ as
\begin{equation}
\frac{\pi L f\sin\varphi_0}{c}=\etaangle,
\quad\Longrightarrow\quad
f=\frac{\etaangle}{\pi}\,\frac{c}{L\,\sin\varphi_0}.
\label{eqn-deltaf-angle}
\end{equation}
The angle-domain \ac{IBW} is twice the one-sided frequency offset
\begin{equation}
\mathrm{IBW}^{\mathrm{ula}}_{a}
=2f
=\frac{2\etaangle}{\pi}\,\frac{c}{L\,\sin\varphi_0}
\approx\frac{0.886\,c}{L\,\sin\varphi_0},
\label{eqn-ula-ibw-angle}
\end{equation}
which depends on the aperture length $L$ and the steering angle $\varphi_0$. It decreases as $L$ increases and as the beam is steered away from the boresight. At boresight ($\varphi_0=0$), it becomes unbounded, consistent with \eqref{eq:squint_angle}, which predicts the absence of angular beam split.

\subsubsection{Distance-Domain IBW}
Retaining the curvature term of \eqref{eqn-ibw-master}, the distance domain gain $g_r(f)$ is given by
\begin{equation}\small
g_r(f) =\left|\frac{1}{N}\sum_{n=1}^{N}
\exp\!\left(j\,\frac{\pi x^2_n\cos^2\varphi_0}{c}\,\frac{f}{r_0}\right)\right|^2 .
\label{eqn-gr-sum}
\end{equation}
\normalsize
Approximating the sum by its integral over $n$ recasts \eqref{eqn-gr-sum} as
\begin{equation}\small
g_r(\gamma)
\approx\frac{\mathcal{C}^2(\gamma)+\mathcal{S}^2(\gamma)}{\gamma^2},
\qquad
\gamma=\frac{L\cos\varphi_0}{2}\sqrt{\frac{2\,f}{c\,r_0}} . 
\label{eqn-gr-fresnel}
\end{equation}
\normalsize
Here, $\mathcal{C}(\cdot)$ and $\mathcal{S}(\cdot)$ denote Fresnel cosine and sine integrals \cite{10443535}. We define $\etarange \triangleq \gamma$ and find $\etarange$ such that $g_r(\etarange)=\rho_\epsilon$, which evaluates to $\etarange \approx1.318$ for $\rho_\epsilon=0.5$. Solving $\etarange = \gamma$ for the two-sided frequency offset yields 
\begin{equation}
\mathrm{IBW}^{\mathrm{ula}}_{r}
=2f
=\frac{4\,\etarange^2\,c\,r_0}{L^2\cos^2\varphi_0}
\approx\frac{7\,c\,r_0}{L^2\cos^2\varphi_0}.
\label{eqn-ula-ibw-range}
\end{equation}
The distance-domain \ac{IBW} increases with the range $r_0$ and with the angle $\varphi_0$ through the factor $\cos^2\varphi_0$. This behavior contrasts with the angle-domain \ac{IBW} in \eqref{eqn-ula-ibw-angle}, which decreases at large angles. The distance-domain \ac{IBW} is linked to radial beam split, a direct consequence of the \ac{NF} effect, which is strongest at boresight and weakest at endfire for a \ac{ULA} \cite{10988573}.

\subsubsection{Combined Near-Field IBW}
The angle-domain and distance-domain \acp{IBW} exhibit opposite trends with respect to angle. At practical distances, one of the two is the limiting factor over most of the angular interval. Consequently, the \ac{NF} \ac{IBW} is approximated by the minimum of the two, given by
\begin{equation}
\mathrm{IBW}^{\mathrm{ula}}
=\min\!\left(\frac{0.886c}{L\sin\varphi_0},\, 
\frac{7c\,r_0}{L^2\cos^2\varphi_0}\right).
\label{eqn-ula-ibw-min}
\end{equation}
The two branches intersect at the crossover distance $r_0 = \frac{L\cos^2\varphi_0}{7.84\sin\varphi_0}$. Radial beam split dominates below this threshold, while angular beam split governs performance above it. The derived expression is accurate in regimes where one effect dominates, but it overestimates the \ac{IBW} near the crossover point where both branches contribute equally.

\begin{figure}[t]
\centering
\resizebox{0.95\columnwidth}{!}{\begin{tikzpicture}[>=stealth, font=\sffamily, line width=0.8pt]

    \coordinate (O)  at (0, 0);          
    \coordinate (Xk) at (1.6, 0);        
    \coordinate (F)  at (4.6, 3.2);      
    
    \draw[->, gray!80] (-4.5, 0) -- (5.3, 0) node[right, text=black] {$x$};
    \draw[->, dashed, gray!80] (0, 0.0) -- (0, 3.0) node[above, text=black] {$y$};
    
    \draw[dashed, blue!60] (1.6, 0) -- (1.6, 3.0);

    \draw[fill=gray!15, rounded corners=1.5pt] (-4, -0.2) rectangle (4, 0.2);
    
    \draw[fill=blue!10] (0.8, -0.2) rectangle (2.4, 0.2);
    
    \draw[thick, black!70] (-2.4, -0.2) -- (-2.4, 0.2);
    \draw[thick, black!70] (-0.8, -0.2) -- (-0.8, 0.2);
    \draw[thick, black!70] (0.8, -0.2) -- (0.8, 0.2);
    \draw[thick, black!70] (2.4, -0.2) -- (2.4, 0.2);
    
    \draw[thick, black!70, rounded corners=1.5pt] (-4, -0.2) rectangle (4, 0.2);
    
    \draw[thick, blue!80!black, rounded corners=1pt] (0.8, -0.2) rectangle (2.4, 0.2);

    \foreach \x in {-3.6, -3.2, -2.8} \filldraw[black!70] (\x, 0) circle (1pt);
    \node at (-3.0, 0.45) {\small $\mathcal{A}_1$};
    
    \foreach \x in {-2.0, -1.6, -1.2} \filldraw[black!70] (\x, 0) circle (1pt);
    \node at (-1.4, 0.45) {\small $\mathcal{A}_2$};
    
    \foreach \x in {-0.4, 0, 0.4} \filldraw[black!70] (\x, 0) circle (1pt);
    \node at (-0.2, 0.45) {\small $\mathcal{A}_3$};
    
    \foreach \x in {1.2, 1.6, 2.0} \filldraw[blue!70!black] (\x, 0) circle (1pt);
    \node[blue!80!black] at (1.4, 0.45) {\small $\mathcal{A}_k$};
    
    \foreach \x in {2.8, 3.2, 3.6} \filldraw[black!70] (\x, 0) circle (1pt);
    \node at (3.0, 0.45) {\small $\mathcal{A}_K$};

    \draw[dashed, gray!80] (-4, -0.2) -- (-4, -0.4);
    \draw[dashed, gray!80] (-2.4, -0.2) -- (-2.4, -0.4);
    \draw[decorate,decoration={brace,amplitude=4pt,mirror}, black!80] 
        (-4, -0.3) -- (-2.4, -0.3) node[midway, below=4pt] {\small $L_k$};

    \filldraw[blue!80!black] (Xk) circle (1.5pt) node[below=7pt] {\small ${x}_k^c$};

    \filldraw[blue] (F) circle (3pt) node[right=4pt, text=black] {\textbf{UE} $(\varphi_0, r_0)$};
    
    \draw[black] (O) -- (F) node[pos=0.6, above left=-3pt] {\small $r_0$};
    \draw[blue!80!black] (Xk) -- (F) node[pos=0.6, below right=-3pt] {\small $r_k$};

    \draw[->] (0, 0.8) arc (90:34.8:0.8) node[midway, above right=-3pt] {\small $\varphi_0$};
    
    \draw[->, blue!80!black] (1.6, 0.8) arc (90:46.8:0.8) node[midway, above right=-3pt] {\small $\varphi_k$};

    \draw[dashed, gray!80] (F) -- (4.6, -0.37);
    \draw[gray!80] (4.4, 0) -- (4.4, 0.2) -- (4.6, 0.2); 

    \draw[dashed, gray!80] (1.6, -0.2) -- (1.6, -0.4);

    \draw[decorate,decoration={brace,amplitude=4pt,mirror}, blue!80!black] 
        (1.6, -0.3) -- (4.6, -0.3) node[midway, below=4pt] {\small $r_0 \sin\varphi_0 - {x}_k^c$};

    \node[text=black!90, align=left] at (-2.5, 2.5) {\small $\varphi_k \neq \varphi_0$};

    \draw[dashed, gray!80] (-1.6, -0.2) -- (-1.6, -0.45);
    \draw[dashed, gray!80] (0, -0.2) -- (0, -0.45);
    \draw[<->, >=stealth, black!80] (-1.6, -0.45) -- node[below=-1pt] {\small $\Delta_s$} (0, -0.45);

    \draw[dashed, gray!80] (-2.0, -0.2) -- (-2.0, -0.45);
    \draw[dashed, gray!80] (-1.6, -0.2) -- (-1.6, -0.45);
    \draw[<->, >=stealth, black!80] (-2.0, -0.45) -- node[below=-1pt] { $\Delta$} (-1.6, -0.45);
\end{tikzpicture}}
\caption{Near-field sub-connected hybrid array, illustrating the local angle $\varphi_k$ and local distance $r_k$ associated with the $\nth{k}$ subarray.}
\label{fig:SC_block_diagram}
\end{figure}
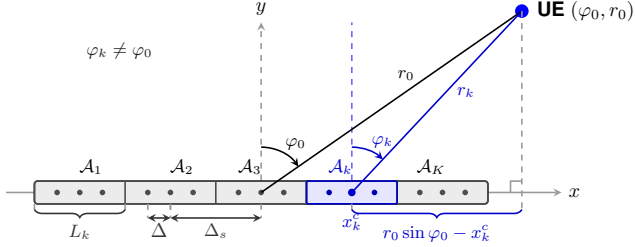

\section{RF Chain Count for Hybrid ULA}\label{sec-IV}
This section derives the required \ac{RF} chain count to mitigate the beam split for \ac{SC} and \ac{FC} hybrid \ac{ULA}. In the \ac{SC} architecture, the block-diagonal analog precoder decomposes the wideband beamforming gain into independent subarray contributions, allowing the required partitioning to be determined directly from the localized \ac{IBW} criterion developed in Section~\ref{subsec:ula_ibw}. In contrast, the dense \ac{FC} analog precoder spans the entire aperture, which instead necessitates an eigenvalue analysis of the space-frequency correlation matrix.

\subsection{Sub-Connected Hybrid ULA}\label{sec-IVA}
The angular and distance \ac{IBW} in \eqref{eqn-ula-ibw-angle} and \eqref{eqn-ula-ibw-range} scale as $1/L$ and $1/L^{2}$, respectively, implying that a shorter aperture supports a wider bandwidth \cite{brown2021active,haupt2015timed}. Partitioning the array into subarrays reduces each subarray's aperture length, thereby increasing its \ac{IBW} and transforming the array-wide bandwidth requirement into a per-subarray \ac{IBW} constraint.

The whole array is partitioned into $K$ contiguous subarrays ${\mathcal{A}}_1,\dots,{\mathcal{A}}_K$, as shown in Fig.~\ref{fig:SC_block_diagram}. Each subarray has aperture length $L_k=L/K$ and comprises $N_k=N/K$ elements, with inter-subarray spacing $\Delta_s=N_k\Delta$. The $\nth{k}$ subarray is centered at ${x}_k^c$ and observes the global focal point $(\varphi_0,r_0)$ at a local angle $\varphi_k$ and range $r_k$. The local range $r_k$ from the focal point to the center of the subarray ${x}_k^c$ is obtained as 
\begin{equation}\label{eq:r_k}
r_k=d({x}_k^c),
\end{equation}
where $d(\cdot)$ is defined in \eqref{eq:fresnel_exact} and $\varphi_k$ can be obtained geometrically from Fig. \ref{fig:SC_block_diagram} as
\begin{equation}\label{eq:phi_k}
\sin\varphi_k=\frac{r_0\sin\varphi_0-{x}_k^c}{r_k}.
\end{equation}
The analog precoder in \eqref{eq:received_signal} for the \ac{SC} hybrid array is the block-diagonal matrix, given by
\begin{equation}\label{eq:precoder_wa}
\mathbf{W}_a=\operatorname{blkdiag}(\mathbf{w}_1,\dots,\mathbf{w}_K)\in\mathbb{C}^{N\times K},
\end{equation}
where $\mathbf{w}_k$ denotes the analog beamforming vector of the $\nth{k}$ subarray. The analog precoder is designed to compensate for the propagation delay to the focal point $(\varphi_0,r_0)$ at the carrier frequency $f_c$, yielding
\begin{equation}\label{eq:precoder_wk}
[\mathbf{W}_a]_{n,k} = \frac{1}{\sqrt{N_k}} e^{-j2\pi f_c\left(d(x_n)-r_0\right)/c},
\quad n\in {\mathcal{A}}_k.
\end{equation}
Since the subarrays occupy disjoint sets of antennas and each beamforming vector has unit norm, the Gram matrix reduces to the identity, $\mathbf{W}_a^{\mathsf{H}}\mathbf{W}_a=\mathbf{I}_K$. Consequently, the digital precoder in \eqref{eq:gf_quot} simplifies to the matched filter $\mathbf{w}_d(f)=\mathbf{W}_a^{\mathsf{H}}\mathbf{a}(f)$. The resulting beamforming gain in \eqref{eq:gf_bound} decomposes into the sum of the individual subarray gains
\begin{align}
g(f) &=\big\|\mathbf{W}_a^{{\mathsf{H}}}\mathbf{a}(f)\big\|^{2}
=\sum_{k=1}^{K} N_k\,g_k(f), \label{eq:gsc}\\
\text{where}, \quad g_k(f)
&=\left|
\frac{1}{N_k}
\sum_{n\in {\mathcal{A}}_k}
e^{-j2\pi f\left(d(x_n)-r_k\right)/c}
\right|^{2}, \label{eq:gkf}
\end{align}
is the normalized gain of the $\nth{k}$ subarray. The digital precoder removes the inter-subarray delay at every frequency. The residual loss across the band is the intra-subarray delay spread carried by $g_k(f)$. The array is split-free over $W$ if every subarray is split-free. The \ac{IBW} criterion is therefore applied to each subarray.
Substituting the local geometry $(L_k,\varphi_k,r_k)$ for $(L,\varphi_0,r_0)$ in \eqref{eqn-ula-ibw-angle} and \eqref{eqn-ula-ibw-range} yields
\begin{equation}
\label{eq:ibwk}
\mathrm{IBW}_k^{\mathrm{ula}}=\min\!\left(\frac{0.886c}{L_k\sin\varphi_k},\ \frac{7\,c\,r_k}{L_k^{2}\cos^{2}\varphi_k}\right).
\end{equation}
The local coordinates $(\varphi_k,r_k)$ vary across subarrays, so the array \ac{IBW} is determined by the binding subarray,
$\displaystyle \mathrm{IBW}=\min_{1\le k\le K}\mathrm{IBW}_k^{\mathrm{ula}}$. Furthermore, a focal point at range $r_0$ lies in the local \ac{FF} of every subarray whenever $r_0\ge\RD/K^{2}$, where $\RD/K^{2}$ is the local Rayleigh distance of each subarray. Assuming a distance of $r_0=L$ yields the requirement $K\ge\sqrt{N}$, which guarantees local \ac{FF} operation for all focal points in
the radiative \ac{NF} of the full aperture. Under this partitioning, the radial beam split mechanism is suppressed within each subarray, and the \ac{IBW} is governed primarily by the angular beam split induced by the distinct local angles $\varphi_{k}$. In this case, the binding subarray is the one experiencing the largest local off-boresight angle, and therefore the smallest angular \ac{IBW}. This per-subarray variation is characteristic of the \ac{NF}. In the \ac{FF}, all subarrays observe the same local angle, so the array \ac{IBW} is equal to that of any subarray.

The array supports a bandwidth $W$ if every subarray satisfies the corresponding \ac{IBW} requirement. Accordingly, the minimum number of subarrays, equivalently the minimum number of \ac{RF} chains, is the smallest partition for which the \ac{IBW} of the binding subarray is no less than $W$
\begin{equation}\label{eq:Krule}
K_\mathrm{SC}^{\mathrm{ula}}=\min\Big\{K'\in\{1,\dots,N\}\ \Big|\ \min_{1\le k\le K'}\mathrm{IBW}_k^{\mathrm{ula}}\ge W\Big\}.
\end{equation}
The \ac{RF} chain count follows by evaluating \eqref{eq:Krule} over the candidate partitions. The stepwise procedure is summarized in Algorithm~\ref{alg:rfcount}.

Unlike \ac{FC} architectures, \ac{SC} hybrid arrays are susceptible to grating lobes if the subarray spacing $\Delta_s$ exceeds half a wavelength \cite{11523528,zhou2026near}. The grating lobes then bound the upper frequency offset $f_c+f_{\mathrm{high}}$, while the beam split bounds the lower one $f_c-f_{\mathrm{low}}$. Grating lobes can be suppressed through non-uniform subarray spacing or overlapped subarrays. Since grating lobes arise from spatial sampling rather than wideband dispersion, we exclude them from the scope of this work, which concerns \ac{NF} beam split analysis.

\begin{algorithm}[t]
\caption{RF Chain Count for the \acs{SC} Hybrid Array}
\label{alg:rfcount}
\renewcommand{\algorithmicrequire}{\textbf{Input:}}
\renewcommand{\algorithmicensure}{\textbf{Output:}}
\begin{algorithmic}[1]
\Require $N$, $L$, $W$, $(r_0,\varphi_0)$
\Ensure Minimum number of \acs{RF} chains $K_\mathrm{SC}^{\mathrm{ula}}$
\For{$K'=1,\dots,N $}
\State Partition the array into $K'$ subarrays
\For{$k=1,\ldots,K'$}
\State Compute $r_k$ and $\varphi_k$ from \eqref{eq:r_k} and \eqref{eq:phi_k}
\State Evaluate $\mathrm{IBW}_k^{\mathrm{ula}}$ from \eqref{eq:ibwk}
\EndFor
\If{$\displaystyle \min_{1\le k\le K'}\mathrm{IBW}_k^{\mathrm{ula}} \ge W$}
\State \Return $K_\mathrm{SC}^{\mathrm{ula}}\gets K'$
\EndIf
\EndFor
\end{algorithmic}
\end{algorithm}

\subsection{Fully-Connected Hybrid ULA}\label{sec-IVB} 
In contrast to the \ac{SC} architecture, each \ac{RF} chain in a \ac{FC} hybrid array is connected to all antenna elements via dedicated \acp{PS}. Consequently, this architecture achieves full aperture gain and yields more \ac{DoF} owing to the additional \acp{PS}. In this subsection, we derive the required \ac{RF} chain count to mitigate the \ac{NF} beam split while achieving the maximum gain. To this end, we analyze the eigenspectrum of the space-frequency correlation matrix, where its \ac{EDoF} dictates the required number of \ac{RF} chains.

\subsubsection{From Average Gain to the Correlation Matrix}
Consider a wideband \ac{NF} beamfocusing at $(\varphi_0,r_0)$ over a bandwidth $W$. We seek the fewest \ac{RF} chain count that holds the focus at $(\varphi_0,r_0)$ across $W$ while attaining the array gain $N$. The average beamforming gain over $W$ in \eqref{eq:gavg} is cast as
\begin{align}
g_{\mathrm{av}}
&\stackrel{(a)}{=}\frac{1}{W}\int_{-W/2}^{W/2}
\big\|(\mathbf W_a^{\mathsf{H}}\mathbf W_a)^{-1/2}\mathbf W_a^{\mathsf{H}}\mathbf a(f)\big\|^2\,df
\label{eq:gavg_b},\\
&\stackrel{(b)}{=}\operatorname{tr}\!\big(\mathbf P_{\mathbf W_a}\mathbf B\big),
\label{eq:gavg_c}
\end{align}
where $(a)$ follows by substituting the maximum beamforming gain in \eqref{eq:gf_bound}, and $(b)$ follows from the identity $\|\mathbf{v}\|^2=\operatorname{tr}(\mathbf{v}\mathbf{v}^{\mathsf{H}})$ \cite[Eq.~(17)]{petersen2008matrix}, together with the cyclic property of the trace, $\operatorname{tr}(\mathbf{AB})=\operatorname{tr}(\mathbf{BA})$ \cite[Eq.~(14)]{petersen2008matrix}. Here $\mathbf P_{\mathbf W_a}=\mathbf W_a(\mathbf W_a^{\mathsf{H}}\mathbf W_a)^{-1}\mathbf W_a^{\mathsf{H}}$ is the orthogonal projection matrix and $\mathbf B \in \mathbb{C}^ {N \times N}$ is the \textit{space-frequency correlation matrix}
\begin{equation}
\mathbf B=\frac{1}{W}\int_{-W/2}^{W/2}\mathbf a(f)\mathbf a^{\mathsf{H}}(f)\,df .
\label{eq:Bmatrix}
\end{equation}
To characterize the \ac{DoF}, we adopt a continuous-aperture model, which is an accurate analytical framework for extremely large-scale arrays \cite{10495774}. In this regime, the discrete antenna locations are replaced by continuous spatial variables $x, x' \in [-L/2, L/2]$. The exact propagation distance from the focal point to an arbitrary point $x$ on the aperture is
\begin{equation}
d(x)=\sqrt{r_0^2+x^2-2r_0x\sin\varphi_0}.
\label{eq:dexact_continuous}
\end{equation}
The discrete spatial correlation matrix $\mathbf B$ is thus generalized to a continuous spatial correlation kernel $B(x,x')$, obtained by integrating the outer product of the continuous array responses over the operating bandwidth
\begin{equation}
B(x,x')=\frac{1}{W}\int_{-W/2}^{W/2}
\exp\!\Big(-j\tfrac{2\pi(f_c+f)}{c}\,\delta(x,x')\Big)\,df,
\label{eq:B_kernel_integral}
\end{equation}
where $\delta(x,x')\triangleq d(x)-d(x')$ is the differential distance between the two aperture points. Evaluating the baseband integral yields
\begin{equation}
B(x,x')= \exp\!\Big(-j\tfrac{2\pi f_c}{c}\,\delta(x,x')\Big) \operatorname{sinc}\!\Big(\tfrac{W}{c}\delta(x,x')\Big).
\label{eq:B_sinc_continuous}
\end{equation}
The eigenspectrum of the spatial correlation operator is governed by the sinc kernel. Because the carrier-dependent exponential term constitutes a unitary phase transformation, it does not alter the eigenvalues of the integral operator \cite{10495774}.

\subsubsection{EDoF of Correlation Matrix}
The discrete spatial correlation matrix admits the eigendecomposition $\mathbf B=\sum_{\ell=1}^{N}\sigma_\ell\bm \nu_\ell\bm \nu_\ell^{{\mathsf{H}}}$, where $\sigma_1\ge\sigma_2\ge\cdots\ge\sigma_N\ge0$, and
$\{\sigma_\ell,\bm \nu_\ell\}_{\ell=1}^{N}$ are the eigenvalue--eigenvector pairs of $\mathbf B$. Substituting this decomposition into \eqref{eq:gavg_c} gives
\begin{equation}
g_{\mathrm{av}}
=\sum_{\ell=1}^{N}
\sigma_\ell
\big\|
\mathbf P_{\mathbf W_a}\bm \nu_\ell
\big\|^{2}.
\label{eq:gavg_eig}
\end{equation}
Since $\operatorname{rank}(\mathbf P_{\mathbf W_a})\le K$, at most $K$ eigenmodes can be captured by the analog beamformer. The average beamforming gain in \eqref{eq:gavg_eig} is maximized when the column space of $\mathbf W_a$ coincides with the eigenspace corresponding to the $K$ largest eigenvalues of $\mathbf B$ \cite{horn2012matrix}, yielding
\begin{equation}
\max_{\operatorname{rank}(\mathbf W_a)\le K}
g_{\mathrm{av}}
=
\sum_{\ell=1}^{K}
\sigma_\ell.
\label{eq:kyfan}
\end{equation}
Therefore, restoring $g_{\mathrm{av}}\rightarrow N$ requires the number of \ac{RF} chains $K$ to equal the number of dominant eigenvalues of $\mathbf B$, defined as the \ac{EDoF}. 

For large $N$, the eigenspectrum of the discrete matrix $\mathbf B$ asymptotically follows that of the continuous spatial kernel $B(x,x')$ defined in \eqref{eq:B_sinc_continuous}. The eigenvalues of this sinc operator exhibit a distinct step-like behavior \cite{7159085}, where the first $\alpha$ eigenvalues capture the entire trace, while the remaining eigenvalues rapidly decay toward zero. Consequently, we have
\begin{equation}
\sum_{\ell=1}^{\alpha}\sigma_\ell
\approx
\operatorname{tr}(\mathbf B)=N, \qquad \sigma_\ell\approx0 \ \ \text{for} \ \ \ell>\alpha,
\label{eq:knee}
\end{equation}
which implies that \ac{EDoF} equals $\alpha$. Therefore, allocating $K_\mathrm{FC}^{\mathrm{ula}}=\lceil\alpha\rceil$ \ac{RF} chains is sufficient to span the dominant eigenmodes and fully restore the array gain. Applying the eigenvalue counting theorem \cite{landau1975szego} to the continuous sinc kernel in \eqref{eq:B_sinc_continuous} shows that the \ac{EDoF} is given by the spatial bandwidth--delay product
\begin{equation}
\alpha = \frac{W}{c} \left( d_{\max} - d_{\min} \right),
\label{eq:alpha}
\end{equation}
where $d_{\max} = \max\limits_{|x| \le L/2} d(x)$ and $d_{\min} = \min\limits_{|x| \le L/2} d(x)$ denote the maximum and minimum path lengths from the focal point to the antenna array, respectively. Their difference, $d_{\max} - d_{\min}$, quantifies the maximum delay spread across the array that induces wideband spatial effects.

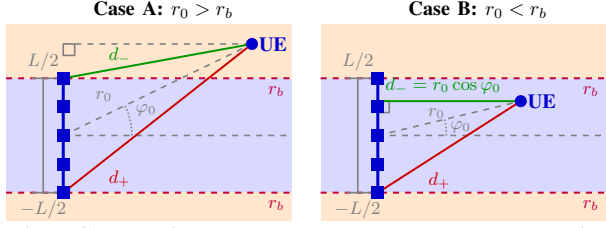
\begin{figure}[t]
\centering
 \resizebox{0.9\columnwidth}{!}{\begin{tikzpicture}[
    scale=1.0, 
    element/.style={fill=blue!80!black, minimum width=6pt, minimum height=6pt, inner sep=0pt}, 
    tauplus/.style={thick, red!80!black, line width=1pt}, 
    tauminus/.style={thick, green!60!black, line width=1pt}, 
    rline/.style={thick, gray, dashed}, 
    boundary/.style={thick, dashed, purple, line width=1pt}
]

\begin{scope}
    \fill[blue!15] (-1.0, -1.0) rectangle (4.0, 1.0);
    \fill[orange!20] (-1.0, 1.0) rectangle (4.0, 1.95); 
    \fill[orange!20] (-1.0, -1.5) rectangle (4.0, -1.0); 
    
    \draw[boundary] (-1.0, 1.0) -- (4.0, 1.0) node[below left, text=purple] {\small $r_b$};
    \draw[boundary] (-1.0, -1.0) -- (4.0, -1.0) node[below left, text=purple] {\small $r_b$};
    
    \draw[gray, dashed, thick] (0,0) -- (4.0,0);
    
    \coordinate (UEA) at (3.3, 1.6);
    
    \draw[thick, gray, dashed] (0, 1.6) -- (UEA);
    \draw[thick, gray, fill=orange!20] (0, 1.4) rectangle (0.2, 1.6);

    \draw[tauplus] (0, -1) -- (UEA) node[pos=0.2, below right, font=\small, text=red!80!black] {$d_{+}$};
    \draw[tauminus] (0, 1) -- (UEA) node[pos=0.2, above right, font=\small, text=green!60!black] {$d_{-}$};
    
    \draw[rline] (0, 0) -- (UEA) node[pos=0.3, above left] {\small $r_0$};
    
    \draw[gray, thick, densely dotted] (1.2, 0) arc (0:25.7:1.2); 
    \node[gray] at (1.45, 0.45) {\small $\varphi_0$};
    
    \draw[gray, thick] (-0.1, -1.0) -- (-0.55, -1.0);
    \draw[gray, thick] (-0.1, 1.0) -- (-0.55, 1.0);
    \draw[gray, thick] (-0.35, -1.0) node[below, font=\small] {$-L/2$} -- (-0.35, 1.0) node[above, font=\small] {$L/2$};

    \draw[line width=1.5pt, blue!80!black] (0, -1.0) -- (0, 1.0);
    \foreach \y in {-1.0, -0.5, 0, 0.5, 1.0} {\node[element] at (0, \y) {};}
    
    \fill[blue!80!black] (UEA) circle (3pt) node[right, text=blue!80!black, font=\bfseries] {UE};
    \node[font=\bfseries] at (1.75, 2.2) {Case A: $r_0 > r_b$};
\end{scope}

\begin{scope}[xshift=5.5cm]
    \fill[blue!15] (-1.0, -1.0) rectangle (4.0, 1.0);
    \fill[orange!20] (-1.0, 1.0) rectangle (4.0, 1.95); 
    \fill[orange!20] (-1.0, -1.5) rectangle (4.0, -1.0); 
    
    \draw[boundary] (-1.0, 1.0) -- (4.0, 1.0) node[below left, text=purple] {\small $r_b$};
    \draw[boundary] (-1.0, -1.0) -- (4.0, -1.0) node[below left, text=purple] {\small $r_b$};

    \draw[gray, dashed, thick] (0,0) -- (4.0,0);
    
    \coordinate (UEB) at (2.5, 0.6);
    
    \draw[thick, gray, fill=blue!15] (0, 0.4) rectangle (0.2, 0.6);
    \draw[tauminus] (0, 0.6) -- (UEB) node[pos=0, above right, font=\small, text=green!60!black] {$d_{-}=r_0\cos\varphi_0$};

    \draw[tauplus] (0, -1) -- (UEB) node[pos=0.3, below right, font=\small, text=red!80!black] {$d_{+}$};
    \draw[rline] (0, 0) -- (UEB) node[pos=0.3, above right] {\small $r_0$};
    
    \draw[gray, thick, densely dotted] (1.2, 0) arc (0:13.5:1.2); 
    \node[gray] at (1.45, 0.15) {\small $\varphi_0$};

    \draw[gray, thick] (-0.1, -1.0) -- (-0.55, -1.0);
    \draw[gray, thick] (-0.1, 1.0) -- (-0.55, 1.0);
    \draw[gray, thick] (-0.35, -1.0) node[below, font=\small] {$-L/2$} -- (-0.35, 1.0) node[above, font=\small] {$L/2$};

    \draw[line width=1.5pt, blue!80!black] (0, -1.0) -- (0, 1.0);
    \foreach \y in {-1.0, -0.5, 0, 0.5, 1.0} {\node[element] at (0, \y) {};}
    
    \fill[blue!80!black] (UEB) circle (3pt) node[right, text=blue!80!black, font=\bfseries] {UE};
    \node[font=\bfseries] at (1.75, 2.2) {Case B: $r_0 < r_b$};
\end{scope}

\end{tikzpicture}}
\caption{In Case A, the transverse component $r_0 {\sin\varphi_0}$ exceeds half the aperture length, so the focal projection falls outside the array, whereas in Case B it is smaller than $L/2$ and the projection falls inside.}
\label{fig:ula_delay_cases}
\end{figure}

\subsubsection{Closed-Form RF Chain Count}\label{sec-IVB3} 
The \ac{RF} chain count in \eqref{eq:alpha} scales with the maximum path-length difference across the aperture via the factor $W/c$, where the path-length difference is determined by the longest and shortest propagation paths from the \ac{UE} to the array. The edge path lengths follow by setting $x=\pm L/2$ in \eqref{eq:dexact_continuous}, as given by
\begin{equation}
d_{\pm}\triangleq
\sqrt{r_0^2+\frac{L^2}{4}\pm r_0L\sin\varphi_0}.
\label{eq:edge_dist}
\end{equation}
 The far edge always provides the longest path, so the maximum path length is $d_+$. The minimum path length instead depends on the orthogonal projection of the focal point onto the array line, whose foot lies at $x^{\star}=r_0\sin\varphi_0$, and two cases arise according to whether this projection falls on the aperture or beyond its edge, as illustrated in Fig.~\ref{fig:ula_delay_cases}. The transition between the two cases occurs when the projection coincides with the array edge $x=L/2$, which gives the boundary
\begin{equation}
r_{\mathrm b}=\frac{L}{2\sin\varphi_0}.
\label{eq:rb}
\end{equation}
This boundary separates the two cases analyzed next.
\paragraph{Case A (Projection Outside the Aperture)}
This corresponds to $r_0\ge r_{\mathrm b}$, where the orthogonal projection falls outside the aperture. In this case, the minimum and maximum path lengths occur at the aperture edges, yielding
\begin{equation}
\alpha_{\mathrm A} = \frac{W}{c}\big(d_+-d_-\big).
\label{eq:alpha_A}
\end{equation}
\paragraph{Case B (Projection Inside the Aperture)}
This corresponds to $r_0<r_{\mathrm b}$, where the orthogonal projection lies within the aperture. The minimum path length is the perpendicular distance from the \ac{UE} to the array line, $d_{\min}
= r_0\cos\varphi_0$ while the maximum path length remains $d_+$. Therefore,
\begin{equation}
\alpha_{\mathrm B} = \frac{W}{c}\big(d_+-r_0\cos\varphi_0\big).
\label{eq:alpha_B}
\end{equation}
Combining both cases yields
\begin{equation}
\alpha(\varphi_0,r_0) = \frac{W}{c}
\begin{cases}
d_+-d_-,
&
r_0\ge\dfrac{L}{2\sin\varphi_0}, \\
d_+-r_0\cos\varphi_0,
&
r_0<\dfrac{L}{2\sin\varphi_0},
\end{cases}
\label{eq:alpha_cases}
\end{equation}
and the minimum required number of \ac{RF} chains is given by $ K_\mathrm{FC}^{\mathrm{ula}} = \lceil\alpha\rceil.$

\begin{figure}[t]
\centering
\begin{minipage}{0.49\columnwidth}
\centering
\includegraphics[width=\linewidth]{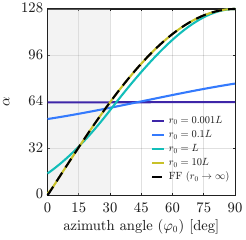}
\end{minipage}\hfill
\begin{minipage}{0.49\columnwidth}
\centering
\includegraphics[width=\linewidth]{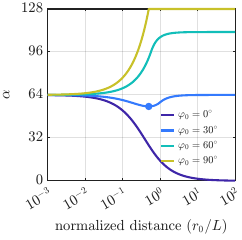}
\end{minipage}
\caption{(Left) $\alpha$ versus azimuth angle $\varphi_0$ for several distances $r_0$. (Right) $\alpha$ versus distance $r_0$ for several azimuth angles.}
\label{fig:ibw_range_angle2}
\end{figure}

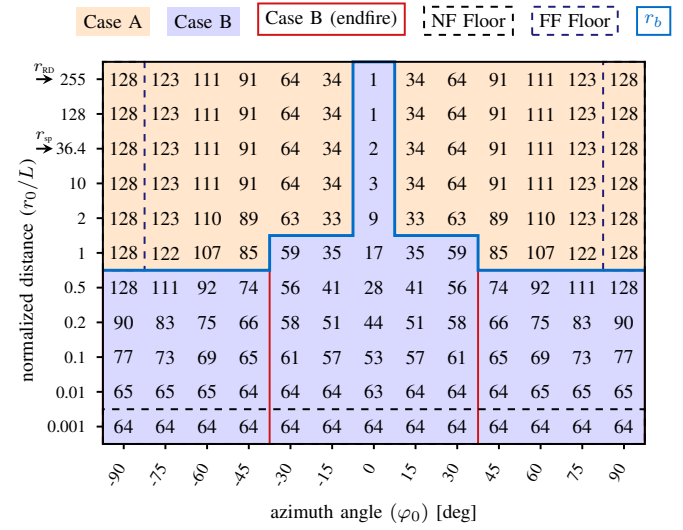
\begin{figure}[t]
\centering
\resizebox{\columnwidth}{!}{\begin{tikzpicture}[x=0.6cm, y=0.5cm]

    \colorlet{caseA}{orange!20}     
    \colorlet{caseB}{blue!15}     
    \definecolor{boxred}{RGB}{210, 20, 20}      
    \definecolor{boxdarkblue}{RGB}{30, 30, 120} 
    \definecolor{boxblue}{RGB}{0, 110, 200}     

    \fill[caseA] (0,0) rectangle (13, 11);
    \fill[caseB] (0,0) rectangle (13, 5);      
    \fill[caseB] (4,5) rectangle (9, 6);       
    \fill[caseB] (6,6) rectangle (7, 11);      

    \tikzset{boundary/.style={line width=0.8pt, line join=miter}}
    
    \draw[boxred, boundary] (0,0) rectangle (4, 5);
    \draw[boxred, boundary] (9,0) rectangle (13, 5);
    
    \draw[boxdarkblue, boundary, dashed] (0,5) rectangle (1, 11);
    \draw[boxdarkblue, boundary, dashed] (12,5) rectangle (13, 11);
    
    \draw[black, boundary, dashed] (0,0) rectangle (13, 1);
    
    \draw[boxblue, line width=1.2pt, line join=miter] 
        (0,5) -- (4,5) -- (4,6) -- (6,6) -- (6,11) -- 
        (7,11) -- (7,6) -- (9,6) -- (9,5) -- (13,5);

    \draw[black, boundary] (0,0) rectangle (13, 11);

    \def\drawrow#1#2{
        \foreach \val [count=\xi] in {#2} {
            \node[font=\footnotesize, anchor=center] at (\xi - 0.5, #1) {\val};
        }
    }
    
    \drawrow{10.5}{128, 123, 111, 91, 64, 34, 1, 34, 64, 91, 111, 123, 128}
    \drawrow{9.5}{128, 123, 111, 91, 64, 34, 1, 34, 64, 91, 111, 123, 128}
    \drawrow{8.5}{128, 123, 111, 91, 64, 34, 2, 34, 64, 91, 111, 123, 128}
    \drawrow{7.5}{128, 123, 111, 91, 64, 34, 3, 34, 64, 91, 111, 123, 128}
    \drawrow{6.5}{128, 123, 110, 89, 63, 33, 9, 33, 63, 89, 110, 123, 128}
    \drawrow{5.5}{128, 122, 107, 85, 59, 35, 17, 35, 59, 85, 107, 122, 128}
    \drawrow{4.5}{128, 111, 92, 74, 56, 41, 28, 41, 56, 74, 92, 111, 128}
    \drawrow{3.5}{90, 83, 75, 66, 58, 51, 44, 51, 58, 66, 75, 83, 90}
    \drawrow{2.5}{77, 73, 69, 65, 61, 57, 53, 57, 61, 65, 69, 73, 77}
    \drawrow{1.5}{65, 65, 65, 64, 64, 64, 63, 64, 64, 64, 65, 65, 65}
    \drawrow{0.5}{64, 64, 64, 64, 64, 64, 64, 64, 64, 64, 64, 64, 64}

    \foreach \x/\val in {0.5/-90, 1.5/-75, 2.5/-60, 3.5/-45, 4.5/-30, 5.5/-15, 6.5/0, 7.5/15, 8.5/30, 9.5/45, 10.5/60, 11.5/75, 12.5/90} {
        \draw[black, thick] (\x, 0) -- (\x, -0.15);
        \node[anchor=east, rotate=60, font=\scriptsize, xshift=-2pt, yshift=-2pt] at (\x, -0.2) {\val};
    }
    \node[font=\footnotesize] at (6.5, -2.0) {azimuth angle $(\varphi_0)$ [deg]};

    \foreach \y/\val in {0.5/0.001, 1.5/0.01, 2.5/0.1, 3.5/0.2, 4.5/0.5, 5.5/1, 6.5/2, 7.5/10, 8.5/36.4, 9.5/128, 10.5/255} {
        \draw[black, thick] (0, \y) -- (-0.15, \y);
        \node[anchor=east, font=\scriptsize] at (-0.2, \y) {\val};
    }
    \node[font=\footnotesize, rotate=90] at (-1.8, 5.0) {normalized distance ($r_0/L$)}; 

    \draw[thick, ->, >=stealth] (-1.6, 10.5) -- (-1.2, 10.5) 
        node[midway, above, font=\scriptsize, inner sep=2pt] {$r_{\mathrm{\scalebox{0.5}{RD}}}$};
        
    \draw[thick, ->, >=stealth] (-1.6,  8.5) -- (-1.2,  8.5) 
        node[midway, above, font=\scriptsize, inner sep=2pt] {$r_{\mathrm{\scalebox{0.5}{sp}}}$};

    \matrix [draw=none, column sep=0.25cm, at={(6.5, 11.5)}, anchor=south, ampersand replacement=\&] {
        \node[fill=caseA, minimum height=0.4cm, inner sep=3pt, font=\footnotesize] {Case A}; \&
        \node[fill=caseB, minimum height=0.4cm, inner sep=3pt, font=\footnotesize] {Case B}; \&
        \node[draw=boxred, line width=0.8pt, fill=white, minimum height=0.4cm, inner sep=3pt, font=\footnotesize] {Case B (endfire)}; \&
        \node[draw=black, line width=0.8pt, dashed, fill=white, minimum height=0.4cm, inner sep=3pt, font=\footnotesize] {NF Floor}; \&
        \node[draw=boxdarkblue, line width=0.8pt, dashed, fill=white, minimum height=0.4cm, inner sep=3pt, font=\footnotesize] {FF Floor}; \&
        \node[draw=boxblue, line width=0.8pt, text=boxblue, fill=white, minimum height=0.4cm, inner sep=3pt, font=\footnotesize] {$r_b$}; \\
    };

\end{tikzpicture}}
\caption{$\alpha$ with respect to angle and distance in the \ac{NF}.}
\label{fig:alpha_range_angle}
\end{figure}

We now illustrate the variation of $\alpha$ in \eqref{eq:alpha_cases} with a numerical example. We consider a \ac{ULA} of $N=256$ half-wavelength-spaced elements, which gives an aperture length $L\approx\unit[1.37]{m}$ at $f_c=\unit[28]{GHz}$. For illustration, we use a deliberately large operating bandwidth $W=\unit[28]{GHz}$. As shown in Fig.~\ref{fig:ibw_range_angle2}, for ${\varphi_0}<30^\circ$, $\alpha$ decreases with increasing $r_0$, whereas for ${\varphi_0}>30^\circ$ it increases with $r_0$. \textbf{In both panels of Fig. \ref{fig:ibw_range_angle2}, the variation becomes negligible beyond approximately $r_0\approx10L$}. 
To further explain the trends, we plot $\alpha$ as a function of distance and angle in Fig. \ref{fig:alpha_range_angle}, with Case A shaded orange and Case B shaded blue, separated by the boundary $r_b$ in \eqref{eq:rb}.

In Case A, $\alpha$ rises with distance and saturates near $r_0=10L$, with its largest value at endfire. Expanding \eqref{eq:edge_dist} for $r_0\gg L$ gives $d_\pm\to r_0\pm\tfrac{L}{2}\sin\varphi_0$, so the delay spread tends to $d_+-d_-\to L\sin\varphi_0$ and the count approaches the \ac{FF} value in \cite{10495774}
\begin{equation}
\alpha_{\mathrm A}\xrightarrow[r_0\gg L]{}\frac{WL\sin\varphi_0}{c}.
\label{eq:alpha_ff}
\end{equation}
At $\varphi_0=90^\circ$, it attains its maximum $WL/c\approx128$, which is the \ac{FF} endfire value.

At boresight, the angular beam split vanishes, and the count reduces to $\alpha=\tfrac{W}{c}\big(\sqrt{r_0^2+L^2/4}-r_0\big)$, which decays as $WL^2/(8cr_0)$ for $r_0\gg L$. The count therefore decreases monotonically with distance until it reaches one at the beam split boundary $r_\mathrm{sp}$ in \eqref{eq:nfbb}. For the very large bandwidth assumed in this example, the \ac{RF} chain count varies with distance even beyond $r_0=10L$. However, for nominal bandwidth values, the \ac{RF} chain count reduces to one well before the beam split boundary. In that case, the \ac{RF} chain count stays nearly constant beyond $10L$, as is the case for the other angles. 


In Case B, $\alpha$ exhibits a lower bound and two characteristic angular regions. For every angle, as $r_0\to0$, the focal point collapses onto the array. Consequently, $d_{\max}\to L/2$ and $d_{\min}\to0$, causing the count to approach an angle-independent \ac{NF} floor
\begin{equation}
\alpha_{\mathrm B}\xrightarrow[r_0\to0]{}\frac{WL}{2c}\approx64,
\label{eq:alpha_floor}
\end{equation}
which is half of the \ac{FF} endfire value. Furthermore, at $\varphi_0=30^\circ$, the \ac{NF} floor $WL/2c$ and the \ac{FF} value $WL\sin\varphi_0/c$ are identical. For $\varphi_0<30^\circ$ the \ac{FF} value lies below the \ac{NF} floor, so the count is higher near the array and lower far away. For $\varphi_0>30^\circ$ it lies above the \ac{NF} floor, so the count is lower near the array and higher far away.

The \ac{RF} chain count is bounded above by the \ac{FF} endfire value $\alpha_{\max}=WL/c\approx128$ and, within the focusing region, below by the angle-independent \ac{NF} floor $WL/2c\approx64$ reached as $r_0\to0$. Since the endpoint crossing at $\varphi_0=30^\circ$ places the \ac{FF} count above this floor for all $\varphi_0>30^\circ$, {a hybrid array dimensioned for \ac{FF} coverage over $|\varphi_0|\ge30^\circ$ already upper bounds the \ac{NF} demand across the focusing region, so no additional \ac{RF} chains are required for \ac{NF} operation}. 


The \ac{RF} chain count for the \ac{FC} architecture can be evaluated numerically from the space-frequency correlation matrix $\mathbf B$ in \eqref{eq:Bmatrix}. The number of dominant eigenvalues of $\mathbf B$, identified by the step behavior in \eqref{eq:knee}, gives $K_\mathrm{FC}^{\mathrm{ula}}=\lceil\alpha\rceil$, and the analog precoder $\mathbf W_a$ is formed from the associated leading eigenvectors $\{\bm \nu_\ell\}_{\ell=1}^{K_\mathrm{FC}^{\mathrm{ula}}}$. Arrays that implement the analog stage with \acp{PS} and no amplitude control restrict $\mathbf W_a$ to entries of constant modulus. The leading eigenvectors are mapped onto this constraint set by alternating projection between the dominant eigenspace and the constant-modulus manifold.

Applying the \ac{IBW} criterion developed in Section~\ref{subsec:ula_ibw} to the \ac{FC} architecture provides a coarse estimate of the required \ac{RF} chain count. Since each \ac{RF} chain connects to all $N$ antenna elements, the effective aperture equals the full array length $L$, allowing \eqref{eqn-ula-ibw-min} to apply directly. Consequently, supporting a total bandwidth $W$ requires $K_{\mathrm{FC}}^{\mathrm{ula}}=\lceil W/\mathrm{IBW}^{\mathrm{ula}}\rceil$ \ac{RF} chains. This expression broadly captures the scaling behavior across all operating regimes. However, minor deviations occur because the \ac{IBW} is defined for an allowable gain-loss threshold ($\rho_\epsilon$), whereas the $\alpha$-based chain count derived in \eqref{eq:alpha_cases} completely mitigates beam split to restore the full beamforming gain.

\section{Extension to a Uniform Circular Array} \label{sec-V}
We now extend the preceding analysis to a \ac{UCA} of radius $R$ and aperture length $L=2R$. The position of each antenna element is represented as $(R, \psi_n)$, where $\psi_n = \frac{2\pi n}{N}$. The \ac{UE} is located at a distance $r$ from the array center, in the direction $(\theta,\varphi)$, where $\theta$ is the elevation angle \cite[Fig.~1]{11498324}. The path length from the \ac{UE} to the $\nth{n}$ antenna element is 
\begin{subequations}
\begin{align}
&d(\psi_n) = \sqrt{r^2+R^2-2Rr\sin\theta\cos(\psi_n-\varphi)}, \label{eq:uca_profile} \\
 &{\overset{(a)}{\approx}} r-R\sin\theta\cos(\psi_n-\varphi) + \frac{R^2}{2r}\left(1-\sin^2\theta\cos^2(\psi_n-\varphi)\right), 
 \label{eq:uca_fresnel}
\end{align}
\end{subequations}\normalsize
where (a) follows from the Fresnel expansion \cite{10988573,11498324,10934779}. 

\subsection{Beam Split}
Following the same approach as in Section \ref{sec-III}, we evaluate the gain on the space--frequency plane using \eqref{eq:g_analog}. Substituting \eqref{eq:uca_fresnel} into \eqref{eq:g_analog} yields the per-element phase mismatch as
\begin{equation}\footnotesize
\begin{split}
\Xi_n(f)=
\underbrace{\frac{R^2}{2}\!\left(f_c\frac{\sin^2\theta_0\cos^2(\psi_n-\varphi_0)}{r_0}
-(f_c+f)\frac{\sin^2\theta\cos^2(\psi_n-\varphi)}{r}\right)}_{\text{curvature (distance) term}} \\ 
-\underbrace{R\big(f_c\sin\theta_0\cos(\psi_n-\varphi_0)-(f_c+f)\sin\theta\cos(\psi_n-\varphi)\big)}_{\text{tilt (angle) term}} .
\label{eq:Xi_uca}
\end{split}
\end{equation}\normalsize
To evaluate the beam split, the tilt and curvature terms are set to zero, yielding the expressions for the angular and radial beam split, respectively.
\paragraph{Angular Beam Split}
Expanding $\cos(\psi_n-\varphi)=\cos\varphi\cos\psi_n+\sin\varphi\sin\psi_n$ in the tilt term of \eqref{eq:Xi_uca} and matching the $\cos\psi_n$ and $\sin\psi_n$ coefficients yields
\begin{equation}
\varphi=\varphi_0,
\qquad
\sin\theta=\frac{1}{1+\bar{f}}\,\sin\theta_0 .
\label{eq:uca_squint_angle}
\end{equation}
Mirroring the \ac{ULA} relation in \eqref{eq:squint_angle}, the elevation angle drifts with frequency. In contrast, the azimuth peak is frequency-invariant. The rotational symmetry of the \ac{UCA} renders an azimuth shift a relabeling of the elements, so no beam split occurs in azimuth.
\paragraph{Radial Beam Split}
Using $\cos^2(\psi_n-\varphi)=\tfrac12\big[1+\cos2(\varphi-\psi_n)\big]$ in the curvature term of \eqref{eq:Xi_uca} and matching the $\cos2\psi_n$ and $\sin2\psi_n$ coefficients yields
\begin{equation}
r=r_0 {(1+ \bar{f})}\,\frac{\sin^2\theta}{\sin^2\theta_0} .
\label{eq:uca_split_match}
\end{equation}
Substituting the split angle $\sin \theta$ from \eqref{eq:uca_squint_angle} simplifies it to
\begin{equation}
r=\frac{r_0}{1+\bar{f}}.
\label{eq:uca_split_range}
\end{equation}
In contrast to the \ac{ULA}, for $f>0$ the \ac{UCA} focal point drifts inward toward the array ($r<r_0$), and outward ($r>r_0$) for $f<0$. Furthermore, this radial split is independent of both the azimuth and elevation angles. 

Following the \ac{ULA} analysis and using the beamdepth of the \ac{UCA} at $f_c+f$, the radial beam split boundary can be shown to coincide with the \ac{EBRD} of the \ac{UCA}, given in \cite[Eq.~(10)]{11498324}. The derivation follows the same steps and is omitted for brevity.

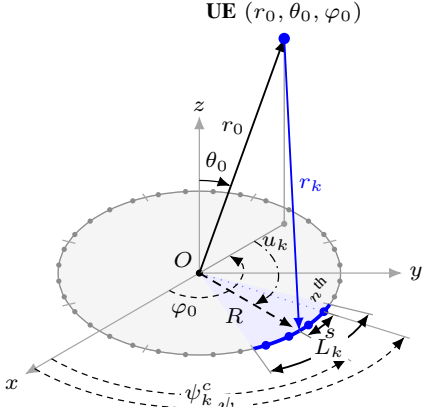
\begin{figure}[t]
\centering
\resizebox{0.65\columnwidth}{!}{
\begin{tikzpicture}[
  font=\scriptsize,
  >={Latex[length=1.5mm,width=1.2mm]},
  axs/.style ={->,thin,black!35},
  r0l/.style ={->,semithick,black},
  rkl/.style ={->,semithick,blue},
  Rl/.style  ={->,semithick,black,dash pattern=on 3pt off 2pt},
  prj/.style ={thin,black!40},
  enl/.style ={thin,blue!55,dotted},
  ath/.style ={->,thin,black},
  aph/.style ={->,thin,black,dash pattern=on 2pt off 1.6pt},
  auk/.style ={->,thin,black,dash pattern=on 3pt off 1.5pt on 0.6pt off 1.5pt},
  apc/.style ={->,thin,blue}, 
  apn/.style ={->,thin,black,dash pattern=on 2pt off 1.6pt}, 
  el/.style  ={circle,fill=black!45,inner sep=0.6pt},
  elk/.style ={circle,fill=blue,inner sep=1.0pt},
  lb/.style  ={inner sep=1pt}
]

\def\Rd{1.60}   \def\ff{0.58}
\def\phio{135}  \def\psic{45}    \def\psinn{61.875}
\def\alo{22.5}  \def\ahi{67.5}
\def\Fpf{0.85}  \def\lift{2.10}
\def\angX{-45} 

\draw[black!16,fill=black!4] (0,0) ellipse ({\Rd} and {\ff*\Rd});
\fill[blue!7] (0,0) --
   plot[domain=\alo:\ahi,samples=40,smooth]
   ({\Rd*sin(\x)},{-\ff*\Rd*cos(\x)}) -- cycle;
\draw[black!45] (0,0) ellipse ({\Rd} and {\ff*\Rd});
\foreach \b in {22.5,67.5,112.5,157.5,202.5,247.5,292.5,337.5}{
  \draw[black!35] ({0.95*\Rd*sin(\b)},{-0.95*\ff*\Rd*cos(\b)})
               -- ({1.05*\Rd*sin(\b)},{-1.05*\ff*\Rd*cos(\b)});}

\foreach \m in {0,...,31}{
  \pgfmathsetmacro\a{5.625+11.25*\m}
  \node[el] at ({\Rd*sin(\a)},{-\ff*\Rd*cos(\a)}){};}
\draw[blue,line width=1.2pt]
   plot[domain=\alo:\ahi,samples=40,smooth]
   ({\Rd*sin(\x)},{-\ff*\Rd*cos(\x)});
\foreach \a in {28.125,39.375,50.625}{
  \node[elk] at ({\Rd*sin(\a)},{-\ff*\Rd*cos(\a)}){};}
\node[elk] (En) at ({\Rd*sin(\psinn)},{-\ff*\Rd*cos(\psinn)}){};

\node[lb, anchor=south east, inner sep=1pt, rotate=45] at ($(En)+(0.16,0.3)$) {$\scriptstyle \nth{n}$};

\coordinate (O)  at (0,0);
\coordinate (P)  at ({\Rd*sin(\psic)},{-\ff*\Rd*cos(\psic)});
\coordinate (Fp) at ({\Fpf*\Rd*sin(\phio)},{-\Fpf*\ff*\Rd*cos(\phio)});
\coordinate (F)  at ($(Fp)+(0,\lift)$);

\draw[axs] (O) -- (0,1.78);              
\node[lb,anchor=south] at (0,1.80) {$z$};
\draw[axs] (O) -- (2.30,0);              
\node[lb,anchor=west]  at (2.34,0) {$y$};
\draw[axs] (O) -- ({1.75*\Rd*sin(\angX)},{-1.75*\ff*\Rd*cos(\angX)});    
\node[lb,anchor=north east] at ({1.77*\Rd*sin(\angX)},{-1.77*\ff*\Rd*cos(\angX)}) {$x$};

\node[circle,fill=black,inner sep=0.8pt] at (O){};
\node[lb,anchor=south east] at (-0.03,0.03) {$O$};

\draw[prj] (F) -- (Fp);
\draw[prj] (O) -- (Fp);
\node[circle,fill=black!40,inner sep=0.7pt] at (Fp){};
\draw[enl] (O) -- (En);

\draw[r0l] (O) -- (F);
\node[lb,anchor=east] at (0.56,1.65) {$r_0$};
\draw[Rl]  (O) -- ($(O)!0.94!(P)$);
\node[lb,anchor=north east] at (0.54,-0.36) {$R$};

\draw[rkl] (F) -- (P) node[pos=0.5,anchor=west,text=blue,inner sep=2pt] {$r_k$};

\node[circle,fill=blue,inner sep=1.3pt] at (F){};
\node[lb,anchor=south] at ($(F)+(0,0.13)$)
      {\textbf{UE} $(r_0,\theta_0,\varphi_0)$};

\draw[ath] (0,1.05) arc (90:70.1:1.05);
\node[lb,anchor=south] at (0.20,1.12) {$\theta_0$};

\draw[aph] ({0.30*\Rd*sin(\angX)},{-0.30*\ff*\Rd*cos(\angX)}) arc(\angX-90:\phio-90:{0.30*\Rd} and {0.30*\ff*\Rd});
\node[lb,anchor=north west] at ({0.40*\Rd*sin(-35)},{-0.40*\ff*\Rd*cos(-35)}) {$\varphi_0$};

\draw[auk] ({0.55*\Rd*sin(\phio)},{-0.55*\ff*\Rd*cos(\phio)}) arc(\phio-90:\psic-90:{0.55*\Rd} and {0.55*\ff*\Rd});
\node[lb,anchor=south west] at ({0.48*\Rd*sin(120)},{-0.48*\ff*\Rd*cos(120)}) {$u_k$};

\draw[black!35,thin] ({1.12*\Rd*sin(\psic)},{-1.12*\ff*\Rd*cos(\psic)})
             -- ({1.52*\Rd*sin(\psic)},{-1.52*\ff*\Rd*cos(\psic)});
\draw[black!35,thin] ({1.12*\Rd*sin(\psinn)},{-1.12*\ff*\Rd*cos(\psinn)})
             -- ({1.69*\Rd*sin(\psinn)},{-1.69*\ff*\Rd*cos(\psinn)});

\draw[apn] ({1.48*\Rd*sin(\angX)},{-1.48*\ff*\Rd*cos(\angX)}) arc(\angX-90:\psic-90:{1.48*\Rd} and {1.48*\ff*\Rd});
\node[lb,fill=white,inner sep=1pt,text=black] at ({1.48*\Rd*sin(0)},{-1.48*\ff*\Rd*cos(0)}) {$\psi_k^{c}$};

\draw[apn] ({1.65*\Rd*sin(\angX)},{-1.65*\ff*\Rd*cos(\angX)}) arc(\angX-90:\psinn-90:{1.65*\Rd} and {1.65*\ff*\Rd});
\node[lb,fill=white,inner sep=1pt,text=black] at ({1.65*\Rd*sin(8.4375)},{-1.65*\ff*\Rd*cos(8.4375)}) {$\psi_n$};

\draw[black!40] ({1.00*\Rd*sin(\alo)},{-1.00*\ff*\Rd*cos(\alo)})
             -- ({1.33*\Rd*sin(\alo)},{-1.33*\ff*\Rd*cos(\alo)});
\draw[black!40] ({1.00*\Rd*sin(\ahi)},{-1.00*\ff*\Rd*cos(\ahi)})
             -- ({1.33*\Rd*sin(\ahi)},{-1.33*\ff*\Rd*cos(\ahi)});
\draw[<->,semithick,black]
   ({1.30*\Rd*sin(\alo)},{-1.30*\ff*\Rd*cos(\alo)})
   arc(\alo-90:\ahi-90:{1.30*\Rd} and {1.30*\ff*\Rd})
   node[midway,fill=white,inner sep=1pt,text=black] {$L_k$};

\draw[black!55] ({1.00*\Rd*sin(\psic)},{-1.00*\ff*\Rd*cos(\psic)})
             -- ({1.12*\Rd*sin(\psic)},{-1.12*\ff*\Rd*cos(\psic)});
\draw[black!55] ({1.00*\Rd*sin(\psinn)},{-1.00*\ff*\Rd*cos(\psinn)})
             -- ({1.12*\Rd*sin(\psinn)},{-1.12*\ff*\Rd*cos(\psinn)});
\draw[<->,semithick,black]
   ({1.08*\Rd*sin(\psic)},{-1.08*\ff*\Rd*cos(\psic)})
   arc(\psic-90:\psinn-90:{1.08*\Rd} and {1.08*\ff*\Rd})
   node[midway,anchor=north west,inner sep=0.5pt] {$s$};

\end{tikzpicture}}
\caption{A \ac{SC} \ac{UCA}, where the $\nth{k}$ subarray observes the \ac{UE} with azimuth separation $u_k=\psi_k^{c}-\varphi_0$ and local range $r_k$.}
\label{fig:uca_sc_geom}
\end{figure}

\subsection{Instantaneous Bandwidth}\label{subsec:uca_ibw}
We derive \ac{IBW} of the \ac{UCA} following the same procedure as the \ac{ULA} case in Section~\ref{subsec:ula_ibw}. 
\subsubsection{Angle-Domain \ac{IBW}}
Retaining only the tilt term of \eqref{eq:Xi_uca}, the normalized space--frequency gain in \eqref{eqn-ibw-master} reduces to
\begin{equation}\small
g_a(f)
=\left|\frac{1}{N}\sum_{n=1}^{N}
\exp\!\left(j\,\frac{2\pi f}{c}\,R\sin\theta_0\cos\psi_n\right)\right|^2 ,
\label{eq:uca-gtheta-sum}
\end{equation}
\normalsize
which, for large $N$, is the ring average of a linear phase and evaluates to the zeroth-order Bessel form $J_0(\cdot)$ \cite{van2002optimum,10243590}
\begin{equation}\small
g_a(f)
=\left|J_0\!\left(\frac{2\pi f}{c}\,R\sin\theta_0\right)\right|^2 .
\label{eq:uca-gtheta-bessel}
\end{equation}
\normalsize
The angle domain \ac{IBW} is obtained by setting $\eta_\mathrm{uca}=\frac{2\pi f}{c}\,R\sin\theta_0$, where the value of $\eta_\mathrm{uca}$ is obtained as $g_a(\eta_\mathrm{uca})=\rho_{\epsilon}$ for a specific threshold. For the 3dB convention $\rho_\epsilon=0.5$, $\eta_\mathrm{uca}\approx1.126$. Solving for the two-sided frequency offset yields
\begin{equation}
\mathrm{IBW}^{\mathrm{uca}}_{a}
=\frac{c\,\eta_\mathrm{uca}}{\pi R\sin\theta_0} \approx \frac{1.126c\,}{\pi R\sin\theta_0},
\label{eq:uca-ibw-angle}
\end{equation}
which depends only on the elevation angle $\theta_0$ and is independent of the azimuth angle.

\subsubsection{Distance-Domain \ac{IBW}}
Retaining only the curvature term of \eqref{eq:Xi_uca}, the gain in \eqref{eqn-ibw-master} reduces to
\begin{align}\small
g_r(f)
&=\left|\frac{1}{N}\sum_{n=1}^{N}
\exp\!\left(j\,\frac{\pi f}{c}\,\frac{R^2\sin^2\theta_0}{r_0}\,\cos^2\psi_n\right)\right|^2 \notag, \\
&\overset{(a)}{\approx}\left|J_0\!\left(\frac{\pi f\,R^2\sin^2\theta_0}{2\,c\,r_0}\right)\right|^2 ,
\label{eq:uca-gr-bessel}
\end{align}
\normalsize
where (a) is derived similarly to the angle domain gain \cite{11498324,10243590}. Solving for the two-sided frequency offset yields the distance-domain \ac{IBW}
\begin{equation}
\mathrm{IBW}^{\mathrm{uca}}_{r}
=\frac{4\,c\,r_0\,\eta_\mathrm{uca}}{\pi R^2\sin^2\theta_0} \approx \frac{4.52\,c\,r_0\,}{\pi R^2\sin^2\theta_0}.
\label{eq:uca-ibw-range}
\end{equation}
\subsubsection{Combined Near-Field \ac{IBW}}
 The combined \ac{IBW} is the minimum of the two \acp{IBW}, expressed as
\begin{equation}
\mathrm{IBW}^{\mathrm{uca}}
=\min\!\left(\frac{1.126c\,}{\pi R\sin\theta_0},\,
\frac{4.52\,c\,r_0\,}{\pi R^2\sin^2\theta_0}\right).
\label{eq:uca-ibw-min}
\end{equation}
The two terms intersect at the crossover range $r_0=R\sin\theta_0/4$. Below this range, the curvature term dominates, making the \ac{IBW} range-limited. Above it, the tilt term dominates, and the \ac{IBW} is governed by \eqref{eq:uca-ibw-angle}.

\subsection{RF Chain Count for the SC UCA}\label{sec-VB}
As shown in Fig. \ref{fig:uca_sc_geom}, the \ac{UCA} is partitioned into $K$ contiguous arcs of angular width $2\pi/K$, each comprising $N_k=N/K$ elements and spanning the arc length $L_k=2\pi R/K=\pi L/K$. The \ac{IBW} in \eqref{eq:uca-ibw-min} does not apply to a single arc because its Bessel kernel results from the full-ring average $\frac{1}{2\pi}\int_{0}^{2\pi}e^{jz\cos\psi}\mathrm{d}\psi=J_0(z)$ over a closed contour. In contrast, a single arc spans only $2\pi/K$ and has two finite endpoints, resulting in the sinc and Fresnel kernels derived for the \ac{ULA} in Section~\ref{subsec:ula_ibw} \cite{10843386}. Accordingly, we characterize the \ac{IBW} of the \ac{SC} \ac{UCA} using \eqref{eq:ibwk}, with the arc-specific geometric parameters replacing those of the \ac{ULA}.


We define $\psi_k^c$ as the mid-angle of the $\nth{k}$ arc and $s=R(\psi_n-\psi_k^c)$ as the arc-length of the $\nth{n}$ element measured from the mid-angle $\psi_k^c$. Substituting $\psi_n=\psi_k^c+s/R$ into \eqref{eq:uca_profile} recasts the element--\ac{UE} distance in terms of the arc-length $s$ 
\begin{equation}\small
d(s)=\sqrt{r_0^{2}+R^{2}-2Rr_0\sin\theta_0\cos\!\big( u_k+s/R\big)},
\ u_k=\psi_k^c-\varphi_0,
\label{eq:uca_ds}
\end{equation}\normalsize
whose value at the arc center is the local range $r_k=d(0)$. Since the elements of an arc are non-collinear, the local angle and distance pair $(\varphi_k, r_k)$ is insufficient to capture the delay spread across the subarray. Instead, we characterize the delay profile using its slope and curvature, obtained from the first and second derivatives of \eqref{eq:uca_ds} evaluated at $s=0$, yielding
\begin{equation}\small
p_k=\frac{r_0\sin\theta_0\sin u_k}{r_k},
\quad
q_k=\frac{r_0\sin\theta_0\cos u_k}{R\,r_k}
-\frac{\big(r_0\sin\theta_0\sin u_k\big)^{2}}{r_k^{3}}.
\label{eq:uca_pq}
\end{equation}\normalsize
The slope $p_k$ replaces $\sin\varphi_k$ and the curvature $q_k$ replaces $\cos^{2}\varphi_k/r_k$ in \eqref{eq:ibwk}, which gives the per-arc \ac{IBW}
\begin{equation}
\mathrm{IBW}_k^\mathrm{uca}=\min\!\left(
\frac{2\etaangle}{\pi}\frac{c}{L_k\,\abs{p_k}},\
\frac{4\etarange^{2}\,c}{L_k^{2}\,\abs{q_k}}\right).
\label{eq:uca_ibwk}
\end{equation}
The minimum \ac{RF} chain count follows from \eqref{eq:Krule} and Algorithm~\ref{alg:rfcount} is evaluated with
$(L_k,p_k,q_k)$.

\subsection{RF Chain Count for FC UCA}
The distance profile in \eqref{eq:uca_profile} is monotonic in $\cos\psi$, with its minimum at $\psi - \varphi_0=0$ and maximum at $\psi-\varphi_0=\pi$. Both extreme distances lie on the ring for every $r_0$ and $\theta_0$, and therefore, unlike the \ac{ULA}, the \ac{UCA} admits no case split. Writing the extreme distances as
\begin{equation}
d_{\pm}=\sqrt{r_0^2+R^2\pm 2Rr_0\sin\theta_0},
\label{eq:uca_edge}
\end{equation}
the \ac{RF} chain count in \eqref{eq:alpha} reduces to the single closed form
\begin{equation}
\alpha_{\mathrm{uca}}(\theta_0,r_0)=\frac{W}{c}\big(d_+-d_-\big)
=\frac{W}{c}\,\frac{4Rr_0\sin\theta_0}{d_++d_-},
\label{eq:uca_alpha}
\end{equation}
and the minimum \ac{RF} chain count is $K_\mathrm{FC}^{\mathrm{uca}}=\lceil\alpha_{\mathrm{uca}}\rceil$. The two extreme elements are diametrically opposite and separated by the diameter $L=2R$, so the triangle inequality gives $d_+-d_-\le L$. The \ac{UCA} therefore never requires more than $WL/c$ \ac{RF} chains.

\subsubsection{Limiting Cases}
The behavior of $\alpha_{\mathrm{uca}}$ is characterized by four limiting regimes.

\paragraph{FF limit ($r_0\to\infty$)}
As the focal point recedes, the spherical wavefront approaches a planar wavefront and $d_+-d_-\to L\sin\theta_0$. Consequently, \eqref{eq:uca_alpha} reduces to
\begin{equation}
\alpha_{\mathrm{uca}}\xrightarrow[r_0\to\infty]{}\frac{WL\sin\theta_0}{c}.
\label{eq:uca_lim_ff}
\end{equation}

\paragraph{Deep NF limit ($r_0\to0$)}
As the focal point approaches the array center, both extreme path lengths converge to $R$, causing the path length difference to vanish, $\alpha_{\mathrm{uca}}\rightarrow 0$. Unlike the \ac{ULA}, the \ac{UCA}
has no \ac{NF} floor.

\paragraph{Axial boresight ($\theta_0=0$)}
All the antenna elements are equidistant from a focal point located along the axial boresight. Hence, the path-length difference across the aperture vanishes for all ranges, yielding $\alpha_{\mathrm{uca}}\big|_{\theta_0=0}=0$.

\paragraph{Co-planar direction ($\theta_0=90^\circ$)}
When the focus lies in the array plane, the path-length difference reduces to $2r_0$ for $r_0<R$ and saturates at $2R$ for $r_0\ge R$. Hence, the \ac{RF} chain count becomes
\begin{equation}
\alpha_{\mathrm{uca}}\big|_{\theta_0=90^\circ}
=
\frac{2W}{c}
\begin{cases}
r_0, & r_0<R,\\
R, & r_0\ge R,
\end{cases}
\label{eq:uca_lim_inplane}
\end{equation}
which increases linearly with $r_0$ until reaching $WL/c$ at $r_0=R$, after which it remains constant.

\begin{figure}[t]
\centering
\begin{minipage}{0.49\columnwidth}
\centering
\includegraphics[width=\linewidth]{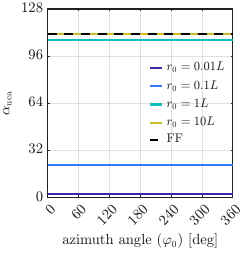}
\end{minipage}\hfill
\begin{minipage}{0.49\columnwidth}
\centering
\includegraphics[width=\linewidth]{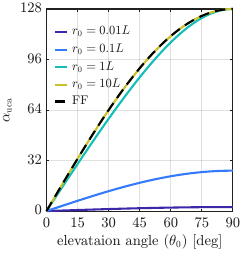}
\end{minipage}
\caption{(Left) $\alpha_{\mathrm{uca}}$ versus azimuth $\varphi$ at $\theta_0=60^{\circ}$ for several distances $r_0$. (Right) $\alpha_{\mathrm{uca}}$ versus elevation $\theta$ at $\varphi_0=0^{\circ}$.}
\label{fig:uca_alpha_cuts}
\end{figure}

Fig.~\ref{fig:uca_alpha_cuts} presents $\alpha_{\mathrm{uca}}$ for a ring of $N=804$ half-wavelength-spaced elements at $f_c=\unit[28]{GHz}$, whose aperture length and operating bandwidth are matched to those of the \ac{ULA} of Fig.~\ref{fig:alpha_range_angle}. The left panel sweeps azimuth at the fixed elevation $\theta_0=60^\circ$, whereas the right panel sweeps elevation for focal distances spanning $0.01L$ to $10L$. Every azimuth cut in the left panel is flat, which confirms that $\alpha_{\mathrm{uca}}$ is invariant to azimuth, as dictated by the rotational symmetry of the ring. In the right panel, all curves originate at $\alpha_{\mathrm{uca}}=0$ for $\theta_0=0$. A focal point on the array axis is equidistant from every ring element, so it induces no differential path length. As the elevation increases, $\alpha_{\mathrm{uca}}$ grows with the in-plane projection $\sin\theta_0$ toward the \ac{FF} envelope. In the coplanar direction, $\alpha_{\mathrm{uca}}$ attains its maximum $WL/c\approx128$. In that plane, $\alpha_{\mathrm{uca}}$ is governed by \eqref{eq:uca_lim_inplane}. Focal points at $r_0>R$ therefore saturate at the diameter-limited value $WL/c\approx128$, whereas points interior to the ring, at $r_0=0.01L$ and $r_0=0.1L$, obey $2Wr_0/c$ and hence require $3$ and $26$ \ac{RF} chains, respectively.

Next, we compare the \ac{UCA} with the \ac{ULA}. In the deep \ac{NF} $\alpha_{\mathrm{uca}}$ approaches zero, whereas $\alpha_{\mathrm{ula}}$ is bounded by the \ac{NF} floor $WL/2c$. In the distance domain, the \ac{UCA} mirrors the \ac{ULA}, with the curves saturating at $r_0\approx10L$.

\section{Numerical Results} \label{sec-VI}
In this section, we provide numerical results to validate the proposed \ac{RF} chain count expressions derived for the \ac{SC} and \ac{FC} hybrid architectures in Sections~\ref{sec-IV} and~\ref{sec-V}. To demonstrate the beam split mitigation, we evaluate the average beamforming gain, defined in \eqref{eq:gavg}, as a function of the number of \ac{RF} chains. Furthermore, we compare the \ac{RF} chain requirements between the \ac{SC} and \ac{FC} architectures across both geometries. Unless specified otherwise, the simulations assume a carrier frequency of $f_c = 28$~GHz and an operating bandwidth of $W = 3$~GHz. The array aperture is fixed at $L = 1.37$~m, which corresponds to $N = 256$ half-wavelength-spaced antennas for the \ac{ULA} and $N = 804$ antennas for the \ac{UCA}.

\begin{figure}[t]
\centering
\includegraphics[width=0.9\columnwidth]{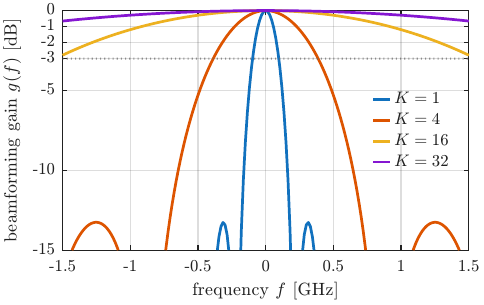}
\caption{Normalized beamforming gain $g(f)$ across the band for a \ac{SC} hybrid \ac{ULA}. We set $\rho_\epsilon=0.5$.}
\label{fig:gain_vs_freq_IBW}
\end{figure}

\begin{figure}
\centering
\includegraphics[width=0.9\columnwidth]{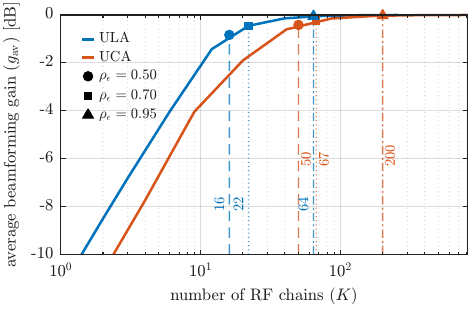}
\caption{Band-averaged \ac{SC} beamforming gain against the number of RF chains $K$ for the \ac{ULA} and the \ac{UCA}.}
\label{fig:gain_vs_K}
\end{figure}

Fig.~\ref{fig:gain_vs_freq_IBW} illustrates the normalized beamforming gain $g(f)$ across the operating band for the \ac{SC} hybrid \ac{ULA}. The evaluation is performed at the focal point $(\varphi_0,r_0)=(90^{\circ},10L)$ for \ac{RF} chain counts $K\in\{1,4,16,32\}$, where Algorithm~\ref{alg:rfcount} returns $K_\mathrm{SC}^{\mathrm{ula}}=16$ for a bandwidth of $W=\unit[3]{GHz}$. As expected from the $1/L_k$ scaling in \eqref{eq:ibwk}, increasing $K$ reduces the physical aperture of each subarray, thereby widening its \ac{IBW}. For the algorithmically predicted count of $K=16$, the maximum in-band loss is $\unit[2.8]{dB}$, occurring at the band edges ($f=\pm\unit[1.5]{GHz}$). Because the threshold defined in \eqref{eq:ibwk} is enforced only at these band edges, the beamforming loss at all inner frequencies remains well below this limit. Moreover, only the binding subarray operates exactly at its \ac{IBW} limit. The remaining subarrays benefit from their specific local geometry $(L_k,\varphi_k,r_k)$ and maintain a wider \ac{IBW}. Consequently, the band-averaged beamforming gain exceeds the worst-case band-edge performance. This is illustrated in Fig.~\ref{fig:gain_vs_K}, which plots $g_{\mathrm{av}}$ versus $K$ for both array geometries. For instance, at $K=16$, the \ac{ULA} achieves an average gain of $g_{\mathrm{av}}=\unit[-1]{dB}$, outperforming the band-edge minimum from Fig.~\ref{fig:gain_vs_freq_IBW} by $\unit[1.8]{dB}$. When comparing geometries, the \ac{UCA} requires approximately $\pi$ times more subarrays than the \ac{ULA}. This occurs because dividing a ring into $K$ segments yields arc lengths of $\pi L/K$, compared to $L/K$ for a linear array, leaving each \ac{UCA} subarray with a larger aperture and a proportionally narrower \ac{IBW}. Finally, the threshold $\rho_\epsilon=0.95$ provides an additional \unit[1]{dB} of gain to achieve the full beamforming gain, but requires four times as many subarrays as the baseline $\rho_\epsilon=0.5$ case. 

\begin{figure}[t]
\centering
\includegraphics[width=0.93\columnwidth]{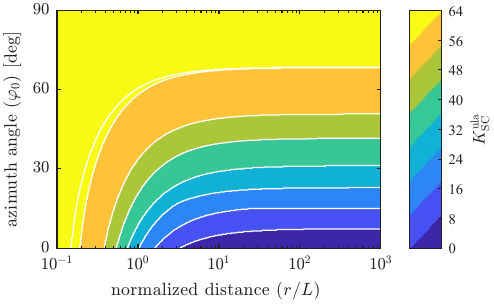}
\caption{Required \ac{RF} chain count for SC hybrid \ac{ULA} over distance and angle. We set $\rho_\epsilon=0.95$.}
\label{fig:RF_chains_SC_FF_vs_NF}
\end{figure}

Fig.~\ref{fig:RF_chains_SC_FF_vs_NF} maps $K_\mathrm{SC}^{\mathrm{ula}}$ over range and angle for $\rho_\epsilon=0.95$. The \ac{RF} chain requirement saturates at a maximum of $64$ along the \ac{FF} endfire direction and also at close distances $r_0\approx0.1L$. This equivalence occurs because both cases place every subarray at a local endfire direction. This behavior is opposite to the \ac{FC} case, where the \ac{NF} floor is half of its \ac{FF} endfire value. However, similar to the \ac{FC} case, the variation of $K_\mathrm{SC}^{\mathrm{ula}}$ saturates beyond approximately $10L$.

\begin{figure}[t]
\centering
\includegraphics[width=0.9\columnwidth]{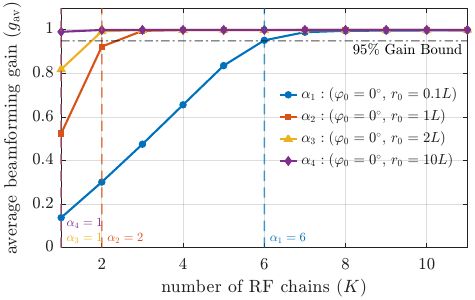}
\caption{The average beamforming gain against the number of \ac{RF} chains for FC hybrid ULA.}
\label{fig:gain_vs_RF_chains}
\end{figure}

\begin{figure}[t]
\centering
\includegraphics[width=0.95\columnwidth]{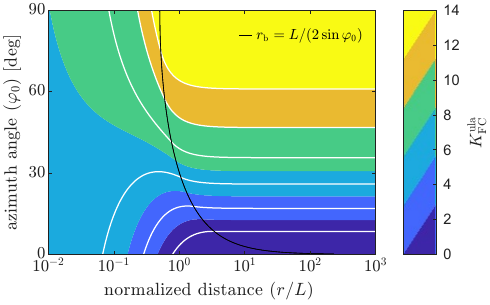}
\caption{Required \ac{RF} chain count for FC hybrid \ac{ULA} over distance and angle.}
\label{fig:RF_chains_FF_vs_NF}
\end{figure}

\begin{figure}
\centering
\includegraphics[height=0.6\columnwidth]{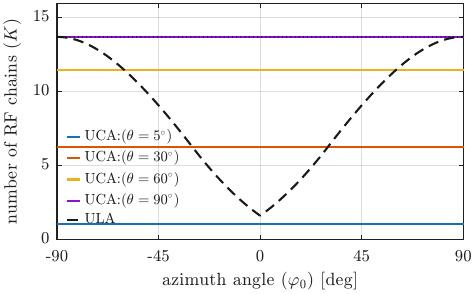}
\caption{\ac{RF} chain count of the \ac{FC} hybrid \ac{ULA} and the \ac{UCA} against azimuth angle.}
\label{fig:RF_chains_ULA_vs_UCA_angle}
\end{figure}

\begin{figure}[t]
\centering
\includegraphics[width=.95\columnwidth]{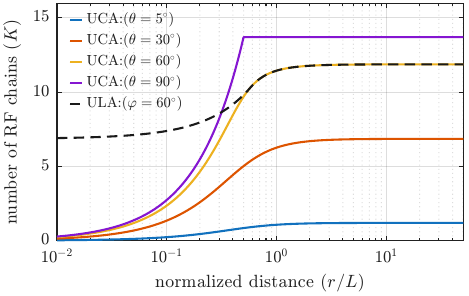}
\caption{\ac{RF} chain count of the \ac{FC} hybrid \ac{ULA} and the \ac{UCA} against normalized distance.}
\label{fig:RF_chains_ULA_vs_UCA_distance}
\end{figure}

\begin{figure}
\centering
\includegraphics[height=0.6\columnwidth]{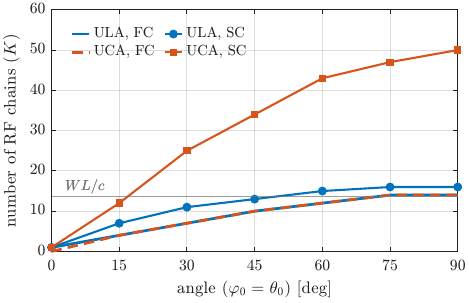}
\caption{\ac{RF} chain count versus angle for the \ac{FC} and \ac{SC} architectures, comparing \ac{ULA} and \ac{UCA}.}
\label{fig:fc_vs_sc}
\end{figure}

Next, we present the results for the \ac{FC} architecture. Fig.~\ref{fig:gain_vs_RF_chains} plots $g_{\mathrm{av}}$ against the \ac{RF} chain count for the \ac{FC} hybrid \ac{ULA} at boresight angle and different ranges. At this angle, the angular beam split vanishes, and the required \ac{RF} chain count is governed entirely by the radial beam split. The four focal ranges $r_0\in\{0.1L,L,2L,10L\}$ yield $\ceil{\alpha}=6,2,1,1$ according to \eqref{eq:alpha_cases}. In each case, the achieved gain crosses the $95\%$ bound exactly at the marker predicted by our analysis. The parameter $\alpha$ decreases as the range increases because the edge-to-center delay spread $d_+-r_0\cos\varphi_0$ decays proportionally to $L^2/(8r_0)$. Note that the bound in \eqref{eq:kyfan} assumes an analog precoder formed from the dominant eigenvectors, which do not satisfy the constant modulus constraint of practical \ac{PS} networks. The precoder used in these simulations is instead obtained via alternating projection onto the constant modulus set. This practical restriction accounts for the residual gap of approximately $5\%$ relative to the unconstrained eigenvector benchmark.

Fig.~\ref{fig:RF_chains_FF_vs_NF} maps $K_\mathrm{FC}^{\mathrm{ula}}$ as a function of both range and angle. The maximum \ac{RF} chain count $WL\sin\varphi_0/c=14$ occurs at the \ac{FF} endfire ($\varphi_0=90^{\circ}$). Within the \ac{NF} region, the boundary $r_b$ defined in \eqref{eq:rb} delineates the two analytical regimes. At extreme \ac{NF} distances ($r_0=0.01L$), the required \ac{RF} chain count drops to exactly half of the \ac{FF} endfire maximum, consistent with the theoretical floor derived in \eqref{eq:alpha_floor}. Conversely, beyond a focal distance of approximately $10L$, the \ac{RF} chain requirement becomes range-invariant. Consequently, assuming $r_0>10L$, neither the \ac{FC} nor the \ac{SC} architecture requires physical reconfiguration of the hardware interconnect when a \ac{UE} transitions between the \ac{NF} and \ac{FF}. However, when the exact \ac{UE} location is known, the transceiver can dynamically power only the required $\lceil\alpha(\varphi_0,r_0)\rceil$ \ac{RF} chains. Leveraging the dynamic hybrid architecture depicted in Fig.~\ref{fig:subarray_types}, this adaptive activation strategy maintains the $95\%$ gain guarantee while improving energy efficiency.

We compare the \ac{RF} chain requirements of the \ac{ULA} and \ac{UCA} geometries in Figs.~\ref{fig:RF_chains_ULA_vs_UCA_angle} and~\ref{fig:RF_chains_ULA_vs_UCA_distance}. Fig.~\ref{fig:RF_chains_ULA_vs_UCA_angle} evaluates the \ac{RF} chain count as a function of the azimuth angle at a fixed range. For the \ac{UCA}, this count remains constant across all azimuth angles, as it is determined only by the elevation angle. Conversely, for the \ac{ULA}, it depends on the azimuth angle. The \ac{RF} chain count reaches a minimum of $2$ chains at boresight (since range is fixed at $r_0=L$) and scales up to $WL/c=14$ at endfire direction. Fig.~\ref{fig:RF_chains_ULA_vs_UCA_distance} plots the required \ac{RF} chains against the normalized distance. Here, the \ac{RF} chain count for the \ac{UCA} rises monotonically from zero and approaches its \ac{FF} limit. In contrast, the \ac{ULA} count starts from a floor of $WL/2c=7$ in the deep \ac{NF}. While the \ac{UCA} requires fewer \ac{RF} chains than the \ac{ULA} at short propagation distances, ranges on the order of the aperture length $L$ are typically encountered only for very large array apertures, limiting the practical significance of this advantage.

Next, Fig.~\ref{fig:fc_vs_sc} compares the \ac{FC} and \ac{SC} architectures in terms of required \ac{RF} chain counts for both array geometries as a function of the angle, evaluated at a fixed distance of $10L$. For a fair comparison, the elevation angle of the \ac{UCA} is set equal to the azimuth angle of the \ac{ULA}. The elevation angle plays the same role in \ac{UCA} as the azimuth angle in \ac{ULA}. The \ac{FC} \ac{RF} chain counts of the two geometries coincide over most of the angular sweep and are bounded by $WL/c$. Conversely, the \ac{SC} counts lie above their \ac{FC} counterparts at every angle. Because the block-diagonal precoder in \ac{SC} systems confines each \ac{RF} chain to a disjoint subset of elements, it cannot span the dominant eigenmodes of $\mathbf{B}$. Instead, it must shrink the physical size of every subarray until its individual \ac{IBW} exceeds $W$. This hardware penalty is modest for the \ac{ULA} but pronounced for the \ac{UCA}, which corroborates the arc-length penalty previously observed in Fig.~\ref{fig:gain_vs_K}. Therefore, to mitigate the beam split, \ac{FC} architectures require fewer \ac{RF} chains than \ac{SC} configurations. However, this reduction in \ac{RF} chains comes at the cost of an increased phase shifter network. Specifically, the \ac{FC} architecture requires ($K \times N$) \acp{PS}, whereas the \ac{SC} architecture requires only $N$ \cite{10495774}. In addition, the larger number of \acp{PS} and the associated RF combining network introduce higher insertion losses, which can offset the performance gains of the \ac{FC} architecture, particularly in wideband beamforming scenarios. 

For a fixed aperture length, a \ac{UCA} accommodates a greater number of antenna elements than a \ac{ULA}. Consequently, in \ac{FC} architectures, the ratio of \ac{RF} chains to antenna elements is lower for the \ac{UCA} than for the \ac{ULA}. Conversely, in \ac{SC} architectures, this normalized \ac{RF} chain count remains approximately equivalent across both array geometries.


Table~\ref{tab:fc_ula} presents the required \ac{RF} chain count $\lceil\alpha\rceil$ for the \ac{FC} \ac{ULA} at $(\varphi_0,r_0)=(90^{\circ},10L)$ across various aperture sizes and bandwidths relevant to 5G and future 6G systems. Because the \ac{RF} chain count scales proportionally with the product $WL$, it doubles when either the aperture or the bandwidth is doubled. For instance, a \unit[1]{m} panel can operate with 7 \ac{RF} chains at \unit[2]{GHz} but requires 14 chains at \unit[4]{GHz}. Likewise, a \unit[2]{m} panel operating at \unit[2]{GHz} demands the same 14 chains. As illustrated by this table, a greater number of \ac{RF} chains will be required as the bandwidth or aperture sizes grow.

\begin{table}
\centering
\caption{RF Chain Count for the FC ULA at $(\varphi_0=90^{\circ}, r_0=10L)$}
\label{tab:fc_ula}
\setlength{\tabcolsep}{9pt}
\renewcommand{\arraystretch}{1.0}
\footnotesize
\begin{tabular}{@{} l cccccccc @{}}
\toprule
 & \multicolumn{8}{c}{$W$~[GHz]} \\
\cmidrule(l){2-9}
$L$~[m] & 0.1 & 0.2 & 0.3 & 0.4 & 1 & 2 & 3 & 4 \\
\midrule
0.5 & 1 & 1 & 1 & 1 & 2 & 4 & 6 & 7 \\
1.0 & 1 & 1 & 2 & 2 & 4 & 7 & 11 & 14 \\
1.5 & 1 & 2 & 2 & 3 & 6 & 11 & 16 & 21 \\
2.0 & 1 & 2 & 3 & 3 & 7 & 14 & 21 & 27 \\
2.5 & 1 & 2 & 3 & 4 & 9 & 17 & 26 & 34 \\
\bottomrule
\end{tabular}
\end{table}

\section{Conclusion} \label{sec-VII}
In this paper, we derived the \ac{RF} chain count required to mitigate the beam split in wideband \ac{NF} hybrid arrays. We characterized the spatial boundary beyond which the radial beam split vanishes and used the \ac{NF} \ac{IBW} to obtain the \ac{RF} chain count for \ac{SC} architectures. For \ac{FC} architectures, the \ac{RF} chain requirement was determined via the eigenspectrum of the space-frequency correlation matrix. We demonstrated that \ac{FC} arrays require fewer \ac{RF} chains than \ac{SC} arrays, trading \ac{RF} chain count for increased phase shifter complexity. Moreover, for the \ac{FC} architecture, \acp{UCA} require fewer \ac{RF} chains than \acp{ULA} in the deep \ac{NF}. This analytical framework provides a foundation for channel estimation via wideband \ac{NF} codebooks constructed from the eigenvectors of the space-frequency correlation matrix \cite{11503877}. Future work will expand this analysis to multipath channels and planar array geometries \cite{11428208}. Additionally, because uniform arc arrays may entail a reduced time-delay spread due to their geometric curvature \cite{10843386}, comparing their wideband performance against \acp{ULA} warrants further investigation.

\bibliographystyle{IEEEtran}
\bibliography{Bibliography/IEEEabrv,Bibliography/my2bib}
\end{document}